\pdfoutput=1
\documentclass[11pt,a4paper]{article}

    \usepackage[T1]{fontenc}
    \usepackage[utf8]{inputenc}
    \usepackage{lmodern}
    \usepackage[margin=2.5cm]{geometry}
    \usepackage{microtype}
    \usepackage{authblk}

    \usepackage[numbers,sort&compress]{natbib}
    \usepackage{url}
    \usepackage{graphicx}%
    \usepackage{multirow}%
    \usepackage{amsmath,amssymb,amsfonts}%
    \usepackage{amsthm}%
    \usepackage{mathrsfs}%
    \usepackage[title]{appendix}%
    \usepackage{xcolor}%
    \usepackage{textcomp}%
    \usepackage{booktabs}        
    \usepackage{algorithm}%
    \usepackage{algorithmicx}%
    \usepackage{algpseudocode}%
    \usepackage{listings}%
    \usepackage[nolist]{acronym}
    \usepackage{comment}
    \usepackage{makecell}
    \usepackage{threeparttable}  
    \usepackage{siunitx}         
    \usepackage{import}

\usepackage[most]{tcolorbox}
\usepackage{xparse}
\usepackage{refcount}

\newcounter{paperbox}

\usepackage{float}

\newfloat{paperboxfloat}{tbp}{lob}
\NewDocumentEnvironment{paperbox}{m}
{
    \refstepcounter{paperbox}

    \begin{paperboxfloat}

    \begin{tcolorbox}[
        enhanced,
        colback=black!5,
        frame hidden,
        width=\linewidth
    ]

    \textbf{Box \thepaperbox:} #1

    \medskip
}
{
    \end{tcolorbox}
    \end{paperboxfloat}
}

    \usepackage[normalem]{ulem}

    \newif\ifshownew
    \shownewfalse 

    \newif\ifshowfinal
    \showfinalfalse  

    \theoremstyle{definition}%

    \newif\ifprintcomments
    \printcommentsfalse
    
    \newcommand{\paul}[1]{\ifprintcomments\textcolor{purple}{[Paul: #1]}\fi}

    \usepackage[left,mathlines]{lineno}
    \newif\ifshowlinenumbers
    \showlinenumbersfalse

    \usepackage[hidelinks]{hyperref}

\newacro{ADC}{analog-to-digital converter}
\newacroplural{ADC}[ADCs]{analog-to-digital converters}

\newacro{MLSA}{match line sense amplifier}

\newacro{DAC}{digital-to-analog converter}
\newacroplural{DAC}[DACs]{digital-to-analog converters}

\newacro{DRAM}{dynamic random-access memory}
\newacroplural{DRAM}[DRAMs]{dynamic random-access memories}

\newacro{DPE}{dot-product engine}

\newacro{RAM}{random-access memory}
\newacroplural{RAM}[RAMs]{random-access memories}

\newacro{TIA}{transimpedance amplifier}
\newacroplural{TIA}[TIAs]{transimpedance amplifiers}

\newacro{SRAM}{static random-access memory}
\newacroplural{SRAM}[SRAMs]{static random-access memories}

\newacro{HBM}{high-bandwidth memory}
\newacroplural{HBM}[HBMs]{high-bandwidth memories}

\newacro{GPU}{graphics processing unit}
\newacroplural{GPU}[GPUs]{graphics processing units}

\newacro{CPU}{central processing unit}

\newacro{TPU}{tensor processing unit}
\newacroplural{TPU}[TPUs]{tensor processing units}

\newacro{IMC}{in-memory computing}
\newacroplural{IMC}[IMCs]{in-memory computing systems}

\newacro{CIM}{compute-in-memory}
\newacroplural{CIM}[CIMs]{compute-in-memory architectures}

\newacro{MAC}{multiply-accumulate}
\newacroplural{MAC}[MACs]{multiply-accumulate operations}

\newacro{VMM}{vector–matrix multiplication}
\newacroplural{VMM}[VMMs]{vector–matrix multiplications}

\newacro{KV}{key-value}
\newacroplural{KV}[KVs]{key-value pairs}

\newacro{LLM}{large language model}
\newacroplural{LLM}[LLMs]{large language models}

\newacro{RNN}{recurrent neural network}
\newacroplural{RNN}[RNNs]{recurrent neural networks}

\newacro{ANN}{artificial neural network}
\newacroplural{ANN}[ANNs]{artificial neural networks}

\newacro{AI}{artificial intelligence}

\newacro{GAI}{generative artificial intelligence}

\newacro{NLP}{natural language processing}

\newacro{PWM}{pulse-width modulation}
\newacroplural{PWM}[PWMs]{pulse-width modulations}

\newacro{QAT}{quantization-aware training}
\newacroplural{QAT}[QATs]{quantization-aware training strategies}

\newacro{ReLU}{rectified linear unit}
\newacroplural{ReLU}[ReLUs]{rectified linear units}

\newacro{CAM}{content-addressable memory}
\newacroplural{CAM}[CAMs]{content-addressable memories}

\newacro{aCAM}{analog content-addressable memory}
\newacroplural{aCAM}[aCAMs]{analog content-addressable memories}

\newacro{ML}{match line}
\newacroplural{ML}[MLs]{match lines}

\newacro{DL}{data line}
\newacroplural{DL}[DLs]{data lines}

\newacro{WL}{word line}
\newacroplural{WL}[WLs]{word lines}

\newacro{BL}{bit line}
\newacroplural{BL}[BLs]{bit lines}

\newacro{WE}{write enable}
\newacroplural{WE}[WEs]{write enable signals}

\newacro{MLSA}{match-line sense amplifier}

\newacro{PCSA}{precharge sense amplifier}

\newacro{PDK}{process design kit}

\newacro{SPICE}{simulation program with integrated circuit emphasis}

\newacro{CMOS}{complementary metal-oxide-semiconductor}

\newacro{MOSFET}{metal-oxide-semiconductor field-effect transistor}
\newacroplural{MOSFET}[MOSFETs]{metal-oxide-semiconductor field-effect transistors}

\newacro{NVM}{non-volatile memory}

\newacro{ReRAM}{resistive random-access memory}

\newacro{MRAM}{magnetoresistive random-access memory}

\newacro{PCM}{phase-change memory}

\newacro{FeRAM}{ferroelectric random-access memory}

\newacro{MTJ}{magnetic tunnel junction}

\newacro{FTJ}{ferroelectric tunnel junction}

\newacro{FeFET}{ferroelectric field-effect transistor}

\newacro{OSFET}{oxide semiconductor field-effect transistor}

\newacro{IGZO}{indium gallium zinc oxide}

\newacro{ITO}{indium tin oxide}

\newacro{HRS}{high-resistance state}

\newacro{LRS}{low-resistance state}

\newacro{SET}{set (programming voltage)}

\newacro{RESET}{reset (programming voltage)}

\newacro{FSM}{finite state machine}
\newacroplural{FSM}[FSMs]{finite state machines}

\newacro{LUT}{lookup table}
\newacroplural{LUT}[LUTs]{lookup tables}

\newacro{VTC}{voltage-to-time converter}
\newacroplural{VTC}[VTCs]{voltage-to-time converters}

\newacro{SDK}{software development kit}
\newacroplural{SDK}[SDKs]{software development kits}

\newacro{IP}{intellectual property}

\newacro{eFLASH}{embedded flash memory}

\newacro{3T1M}{three-transistor one-memristor}

\newacro{6T2M}{six-transistor two-memristor}

\newacro{8T2M}{eight-transistor two-memristor}

\newacro{SALM}{strong-arm latched memristor}

\newacro{LB}{low bound}

\newacro{HB}{high bound}

\newacro{AM}{Associative Memory}
\newacroplural{AM}[AMs]{Associative Memories}

\newacro{GM}{memristor conductance}

\newacro{PE}{processing element}
\newacroplural{PE}[PEs]{processing elements}

\newacro{NOC}{network-on-chip}
\newacroplural{NOC}[NOCs]{networks-on-chip}

\newacro{RISC-V}{reduced instruction set computer version V}
\newacroplural{RISC-V}[RISC-Vs]{reduced instruction set computer version V processors}

\newacro{IRIS}{Iris flower dataset}

\newacro{SME}{small and medium enterprise}
\newacroplural{SME}[SMEs]{small and medium enterprises}

\newacro{BMFTR}{Bundesministerium für Forschung, Technologie und Raumfahrt}

    \title{\textbf{Perspective:} Content-addressable memories as a computing primitive for today's AI and beyond}

    \author[1,2,*]{Paul-Philipp Manea}
    \author[3,4]{Bo Wen}
    \author[3,4]{Peiyi He}
    \author[3,4,*]{Can Li}
    \author[1,2,*]{John Paul Strachan}

    \affil[1]{Peter Gr\"unberg Institute (PGI-14), Forschungszentrum J\"ulich, J\"ulich, Germany}
    \affil[2]{Faculty of Electrical Engineering, RWTH Aachen University, Aachen, Germany}
    \affil[3]{Department of Electrical and Computer Engineering, The University of Hong Kong, Pokfulam, Hong Kong SAR, China}
    \affil[4]{Center for Advanced Semiconductors and Integrated Circuits, The University of Hong Kong, Pokfulam, Hong Kong SAR, China}
    \affil[*]{Corresponding authors: \href{mailto:p.manea@fz-juelich.de}{p.manea@fz-juelich.de}, \href{mailto:canl@hku.hk}{canl@hku.hk}, \href{mailto:j.strachan@fz-juelich.de}{j.strachan@fz-juelich.de}}

    \date{}

\begin{document}
    \ifshowlinenumbers
      \linenumbers
    \fi

    \maketitle

\begin{abstract}
Modern artificial intelligence is predominantly executed on computing architectures optimized for dense linear algebra. While this has enabled the success of contemporary neural networks and motivated \ac{CIM} architectures, a growing class of \ac{AI} workloads depends on associative retrieval, identifying stored information by content or similarity rather than by explicit memory addresses. Such operations are central to transformer attention, tree-based inference, genomic search, and other retrieval-intensive applications, yet remain inefficiently supported by conventional memory systems. In this Perspective, we argue that \acp{CAM} provide a complementary hardware primitive for associative processing in \ac{AI}. We review their ability to perform massively parallel in-memory matching, discuss how emerging memory technologies can improve density and energy efficiency, and identify hierarchical search, application-specific architectures, hardware-aware learning, and heterogeneous integration with \ac{CIM} as key directions for scalable associative computing. Together, these developments suggest that associative retrieval should complement linear algebra as a foundational computing primitive for future \ac{AI} systems.
\end{abstract}

    \section{Introduction}
        \begin{figure}
        \centering
        \includegraphics[width=1\linewidth]{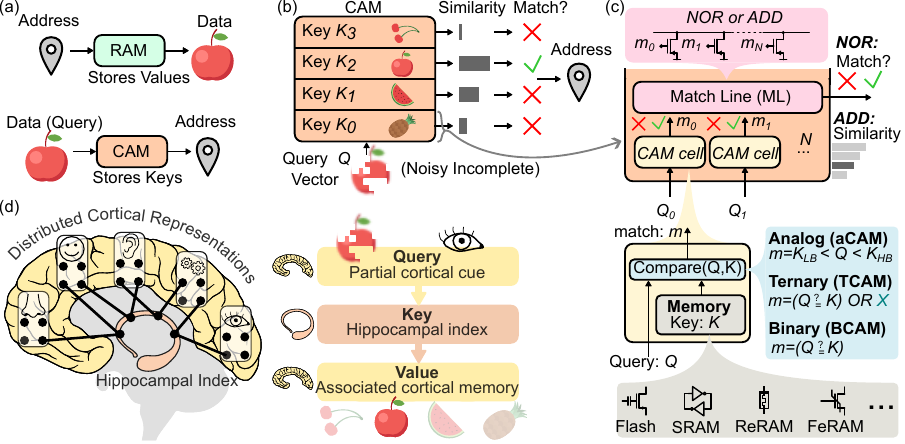}
        \caption{\textbf{Associative memories in artificial and biological memory systems.} (a) Comparison between conventional \ac{RAM} and \acp{CAM}. While \ac{RAM} retrieves data from a supplied address, \acp{CAM} retrieve the address or association corresponding to a supplied query. (b) In a \ac{CAM} array, an input query is compared against many stored keys in parallel, producing a match or similarity score for each entry, which can then be encoded into the address of the exact or closest match. (c) General operating principle of associative search in a \ac{CAM}. Individual cells compare query and stored key values to generate local match signals, which are subsequently combined along a match line to produce either a word-level match through a NOR operation or a similarity score through accumulation. Different \ac{CAM} cell architectures support different comparison modalities, including binary, ternary, and analog matching, and can be implemented using a wide range of memory technologies. (d) Conceptual illustration of hippocampal indexing theory, in which a partial cortical cue initiates retrieval through a sparse hippocampal index. This index acts as a memory key that links the cue to distributed cortical representations containing the associated sensory and semantic information, thereby forming a biological query–key–value memory system.
        }
        \label{fig:cam_intro}
    \end{figure}

    \acresetall
    Traditional computer memories operate according to the \ac{RAM} paradigm, where an input address directly connects to a stored digital value. Content-addressable memories (CAM) invert this relationship: a query pattern is presented to memory and compared against a collection of stored keys, as illustrated in Fig.~\ref{fig:cam_intro}(a,b), with the output being a measure of association, such as a similarity score or an exact match, that can be translated into an address \cite{1599540}. This fundamentally changes the nature of memory operations: retrieval itself becomes a computation, and information can be recalled from data that is incomplete, noisy, or partially corrupted by identifying the closest matching stored pattern \cite{Lee1993InexactCAM,he2024shiftcam}. As we discuss in this Perspective, these properties align closely with many computing applications, including today's AI  where the central operations are often not numerical computation alone, but the retrieval of stored patterns by association. 
    
    Despite the diversity of implementations, all \ac{CAM} hardware shares the same underlying computational principle, illustrated in Fig.~\ref{fig:cam_intro}(c). At its core, every \ac{CAM} consists of two fundamental building blocks: a \ac{CAM} cell that stores a key and evaluates its relationship to an input query, and a match line that combines the resulting cell responses $m_{i,j}$ into a word-level result \cite{1599540}. Depending on the match-line architecture, the output may represent either an exact match or a continuous similarity score \cite{Pedretti2022differentiableCAM}. At the cell level, the comparison operation can range from binary and ternary matching, including a ``don’t-care’’ state $X$, to analog range evaluation \cite{Li2020}, and further to other graded similarity metrics that are aggregated along the match \cite{Pedretti2022differentiableCAM,9106766}, while the storage of the key can be implemented using a wide variety of memory technologies.
    
    Beyond just hardware, associative memories have a long history in both computer engineering and neuroscience, although the two fields developed largely independently for many decades. In computer engineering, specialized \ac{CAM} hardware developed for low-latency content-based search and retrieval, initially targeting networking and database applications (Box~\ref{box:hist_computing}). In contrast, neuroscience regarded associative recall as a central principle of biological memory, leading to influential theories including hippocampal indexing and neural associative memories (Box~\ref{box:hist_brain}), illustrated in Fig.~\ref{fig:cam_intro}(d).

\begin{figure}[htb!]
    \centering
    \def\svgwidth{\textwidth}

    \begingroup
    \fontfamily{phv}\selectfont
    \fontsize{6}{7.2}\selectfont
    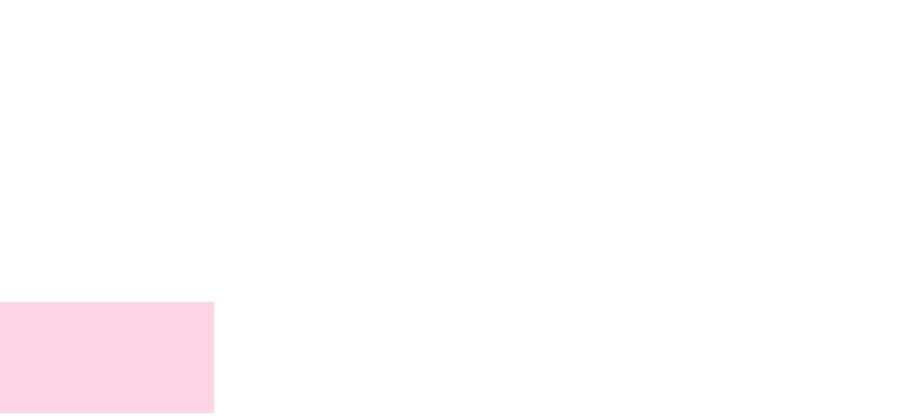
    \endgroup

    \caption{\textbf{Evolution of \acp{CAM} and associative computing.} Timeline of \ac{CAM} and associative-memory development, spanning foundational work on content-based retrieval in the 1960s, the networking era of the 1990s, during which \acp{CAM} became established as practical lookup engines, and their recent emergence as hardware primitives for \ac{AI}, machine learning, and associative computing, enabled by emerging memory technologies and new application domains.}
    \label{fig:CAM_Timeline}
\end{figure}

    \begin{paperbox}{A Brief History of Associative Memory in Computing}
    \label{box:hist_computing}
    The timeline of Fig.~\ref{fig:CAM_Timeline} traces associative memory from its 1960s foundations through networking applications to today’s emerging appliocations in AI.
    
    Early work in the 1960s established key concepts of associative memories such as parallel file search and associative address translation \cite{5219279,Kilburn1961AtlasSupervisor}. Commercial \acp{CAM} emerged in the 1970s, establishing parallel word matching as a practical hardware primitive for search-intensive applications \cite{Intel1973C3104CAM}.

    In the early 1990s, \acp{CAM} became increasingly important for database and networking workloads, where binary and ternary architectures enabled fast lookup, routing, and packet-forwarding operations \cite{34085,mcauley1993fast,zane2003coolcams}. At the same time, research extended \acp{CAM} beyond exact matching toward approximate and similarity-based retrieval \cite{Lee1993InexactCAM}. CMOS VLSI implementations based on SRAM, floating-gate, and DRAM technologies also matured during this period, establishing the practical foundations of integrated \ac{CAM} hardware \cite{34085,130700,Yamagata1991StackedCapacitorCAM}.
        
    More recently, since the 2010s, emerging non-volatile memories, including resistive and ferroelectric devices, have enabled denser and more energy-efficient \ac{CAM} implementations \cite{7063054,5471063,ni2019ferroelectric}. In parallel, \acp{CAM} have expanded beyond networking toward machine learning, nearest-neighbor search, tree inference, differentiable associative memories, and transformer attention \cite{Pedretti2021Tree,Pedretti2022differentiableCAM,9656637,ramsauer2021hopfieldnetworksneed,gao2025sb}.
    \end{paperbox}

    \begin{paperbox}{Associative Memory in the Brain}
    \label{box:hist_brain}
    An addressable and directly programmable RAM-like memory has certainly not been found in the brain. Instead, a multitude of memory types (episodic, semantic, procedural) are described by psychologists and neuroscientists, where memory ``fetching" is instead a dynamic process triggered by context and association (e.g., Proust's madeleine), and memory ``programming" is instead a learning procedure (e.g., learning to ride a bike).  Thus, the view of the brain as an associative memory system has deep roots. In hippocampal indexing theory, the hippocampus stores indices to distributed neocortical activity patterns, linking episodic memories to the cortical areas in which their semantic contents are represented, as shown in Fig.~\ref{fig:cam_intro}(d) \cite{teyler1986hippocampal,Goode2020IntegratedIndex}. Importantly, this does not imply that long-term memory resides only in the hippocampus: cortical networks are thought to support long-lasting semantic memories over years, whereas hippocampal memories are more episodic and gradually consolidated into cortical representations. Recent models extend this idea toward key-value memory, where sparse hippocampal ``keys'' retrieve distributed neocortical ``values'' containing sensory and semantic information \cite{chandra2025episodic,gershman2025key}. Retrieval is likely not an exhaustive global search, but instead relies on sparse and hierarchical hippocampal-cortical dynamics that organize memories into nested subspaces for efficient access to spatial, episodic, and linguistic associations \cite{Peer2025CognitiveMaps,cornell2025hierarchical}. Beyond hippocampal indexing, cortico-thalamic circuits may support attention-like key--value retrieval, where cortical activity maintains short-term memory traces while thalamic pathways contribute to routing, gating, and sustaining these representations \cite{granier2025multiheadselfattentioncorticothalamiccircuits}. This connects biological associative memory to query-key-value mechanisms in transformer architectures and motivates \ac{CAM}-based attention hardware, as discussed in Section~\ref{sec:CAM_attetion}.
    \end{paperbox}

    Modern \ac{AI} increasingly highlights associative retrieval as a central computational principle. The remarkable success of transformer-based \acp{LLM} suggests that attention, interpretable as a form of associative memory, may be fundamental to intelligent computation \cite{vaswani2023attentionneed,ramsauer2021hopfieldnetworksneed,gershman2025key,chandra2025episodic}. This creates a natural bridge between associative memory concepts in neuroscience and content-addressable memory architectures in computer engineering. Recent \ac{CAM}-based implementations of transformer attention further illustrate this convergence \cite{manea2025gaincellcam,molomochir2025camformerassociativememoryneed}.

    The memory bottleneck of conventional von Neumann architectures built around \ac{RAM}, together with the rapidly rising computational and energy demands of \ac{AI}, expose the limits of existing computing systems and the need for more performant, specialized hardware innovations \cite{vogginger2024neuromorphichardwaresustainableai}. This provides a central motivation for reconsidering \acp{CAM} as a missing computational primitive, particularly for workloads in which stored representations must be repeatedly searched, compared, or filtered. Transformer attention is a prominent example, since large volumes of stored key-value representations must be accessed and compared during inference \cite{leroux2025analog}. Although attention can be implemented through linear algebra, this largely reflects the structure of existing GPU and accelerator architectures rather than the most natural form of associative retrieval. Alternative similarity or distance metrics may provide a more suitable computational foundation \cite{Millidge2022UniversalHopfield}. \acp{CAM} address this gap by enabling associative search directly within memory. By supporting distance-based matching, selective nearest-neighbor retrieval, and explicit matching boundaries, they complement conventional linear-algebra \ac{CIM} accelerators and provide a hardware-native substrate for explicit and controllable associative retrieval in \ac{AI} systems.
    
    It is therefore timely to re-evaluate \acp{CAM} beyond their traditional roles in networking and database search. This shift is motivated by a rapidly growing body of work demonstrating that \ac{CAM}-based associative search is a natural computational primitive for an increasingly diverse set of applications. Beyond Transformer attention, \acp{CAM} have been successfully applied to decision-tree inference \cite{Pedretti2021Tree}, bioinformatics \cite{he2024shiftcam}, vector symbolic computing \cite{9921397}, memory-augmented neural networks \cite{Mao2022MemristiveMANN}, and $k$-nearest-neighbor classification and clustering \cite{ni2024tap}. At the same time, advances in memory devices and integration technologies are enabling denser, more scalable, and more energy-efficient \ac{CAM} implementations \cite{5471063,ni2019ferroelectric,liu2021ultrafast,jiang2024scalable,gao2025sb}.

    In this perspective, we argue that \acp{CAM} are undergoing a transition from specialized search engines to a broader computing substrate for associative information processing. We discuss how recent advances in AI, emerging memory technologies, and system architectures are creating new opportunities for \ac{CAM}-based computing, while also introducing important challenges in scalability, search organization, and learning. By connecting developments across devices, circuits, architectures, and applications, we outline a roadmap toward large-scale associative computing systems and explore their potential role in future \ac{AI} hardware.

    The remainder of this paper is organized as follows. We first review the technological foundations of \acp{CAM}, focusing on the memory technologies that enable associative computing. We then survey major \ac{CAM}-based application domains, illustrating their growing importance for future \ac{AI} infrastructures. Finally, we examine \ac{CAM}-based architectures and system-level design considerations, outlining the engineering challenges and opportunities associated with deploying \acp{CAM} across diverse application domains.

    \section{Memory and Device Technology for \ac{CAM} Systems}

    \subsection{Technology Landscape of Associative Memories}
    \label{sec:tech-landscape}
    CMOS SRAM-based T\acp{CAM} are the established industrial standard, providing high-speed, reliable search with mature fabrication \cite{34085} making them well suited for latency-critical applications such as networking, packet classification, and routing. However, their large cell area highly limits capacity. Although both SRAM- and DRAM-based \acp{CAM} store only one bit per memory cell, DRAM-based \acp{CAM} partly address the area limitation through more compact dynamic storage, at the cost of reduced robustness, retention, or circuit simplicity \cite{Yamagata1991StackedCapacitorCAM}.
    
    Flash- and floating-gate-based \acp{CAM} form a dense class of non-volatile associative memories, from early multi-level Si-based floating-gate MOS \acp{CAM} \cite{130700}, to recent 3D-stackable concepts based on thin-film oxide transistor-based such as IGZO and on low-dimensional materials such as MoS$_2$ \cite{10521670,11041521,gao2025sb}. 
    FeFETs instead store the state in the polarization of a ferroelectric gate rather than as charge on a floating gate, avoiding the charge injection or tunneling required for Flash programming and thereby enabling substantially faster, lower-voltage state updates. This gives the same non-volatile, multi-level threshold shift while switching at a few volts in nanoseconds rather than at the tens of volts and microseconds required for tunneling \cite{ni2019ferroelectric,9106766,9656637,9372119,Ni2019}.
    In both FeFET and floating-gate \acp{CAM}, programmable analog thresholds can directly define the matching bounds of an analog \ac{CAM} cell, avoiding additional thresholding circuitry. Their main limitation is that the matching boundaries inherit the finite switching slope and variability of the transistor, which can soften thresholds and limit the number of reliable states or the achievable inference precision, depending on the application.
    
    Memristive technologies, including phase-change and filamentary resistive switching devices, are attractive \ac{CAM} candidates due to their high density, non-volatility, and analog programmability. Early hybrid memristor-CMOS \acp{CAM} demonstrated integration with CMOS search circuitry, followed by many memristive B\ac{CAM} and T\ac{CAM} architectures \cite{5471063,7063054,6858404,7563786,7894184,Masoodur2t2rtcam}. The field later expanded to analog \acp{CAM}, particularly 6T2M cells that store continuous matching boundaries instead of binary states \cite{Li2020}. Since memristors are passive two-terminal devices, CMOS support is typically needed for thresholding and gain, creating a density--discrimination trade-off: additional circuitry sharpens search boundaries but increases cell area and reduces storage density.
    
    Gain-cell-based a\acp{CAM} combine compact dynamic storage with analog associative search and storage functionality \cite{maena_gain_cell_aCAM}. Their high-speed write operation and effective unlimited write endurance make them attractive for associative cache hierarchies, while oxide-semiconductor platforms such as IGZO could enable monolithic 3D integration of dense vertical \ac{CAM} arrays \cite{9401798,10268258,lu2024high,shi_counteractive_2023,10185398,ye_double-gate_2020}.
    
    Additional non-volatile \ac{CAM} concepts based on magnetic tunnel junctions (MTJs) \cite{Matsunaga2012A3U} and, more recently, ferroelectric memcapacitors \cite{Ryu2025}, further broaden the device landscape, although they are currently less developed than FeFET-, Flash-, and memristor-based implementations.
    
    Optical and opto-electronic \acp{CAM} exploit optical parallelism and wavelength-division multiplexing to address electrical bandwidth and interconnect limits \cite{8320858,alkabani2020oe}. They may offer high search throughput for specialized workloads, but large-scale deployment remains limited by fabrication, thermal stability, and opto-electronic integration challenges.
    
    \subsection{Associative Memory Hierarchy}
    \label{sec:memory-hierarchy}

\begin{figure}[htb!]
        \centering
        \includegraphics[width=1\linewidth]{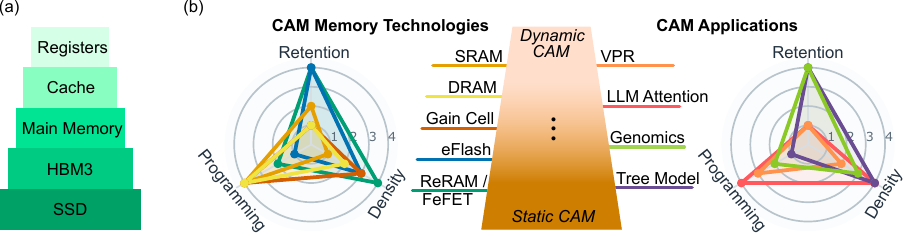}
        \caption{\textbf{From addressed-memory to associative-memory hierarchies.} (a) Conventional addressed-memory hierarchy, represented as a pyramid from registers and caches to main memory, HBM3, and SSDs, reflecting the trade-off between fast, small memories and larger, higher-capacity memories. 
        (b) Envisioned associative \ac{CAM} hierarchy, where representative \ac{CAM} technologies and application classes are arranged along a spectrum from dynamic, write-intensive \ac{CAM} layers to more static, high-capacity \ac{CAM} layers. The radar plots qualitatively compare technologies and workloads in terms of programmability (speed and energy), retention, and density; for gain-cell \acp{CAM}, dense 3D integration is assumed.}
        \label{fig:memory-technology}
    \end{figure}

    Conventional memory is organized as a hierarchy because no single technology simultaneously optimizes density, latency, bandwidth, energy efficiency, and retention, so each level is implemented in a different technology, as shown in Fig.~\ref{fig:memory-technology}(a) \cite{592312}. \ac{CAM} systems will likely follow the same principle. The radar plots of Fig.~\ref{fig:memory-technology}(b) compare candidate technologies (left) with the requirements of representative applications (right) along three axes: density, retention, and programmability, where the latter combines write speed, energy, and endurance. Programmability and density are largely in tension: fast-to-write charge-based cells are either large, as in SRAM, or volatile, as in DRAM and gain cells, whereas the non-volatile mechanisms that provide density and retention require slower, endurance-limited programming. Applications separate along the same axis, since frequently updated associative state favors programmable memories, whereas large but rarely updated search databases prioritize density and retention. Both plots therefore order along a common spectrum, drawn in the center panel.

    For the remainder of this paper we  reduce it to its two endpoints \cite{11040774}: a write-heavy \emph{dynamic} \ac{CAM} tier for frequently updated state, and a larger-capacity \emph{static} \ac{CAM} tier for search-intensive workloads. This two-tier picture is a deliberate simplification, and intermediate tiers are both conceivable and likely.
    
    \paragraph{Tier-1 Dynamic / Write-intensive \acp{CAM}}
    Dynamic \acp{CAM} are high-speed associative memories optimized for frequently updated representations such as embeddings, descriptors or sensory states. Operating close to \ac{AI} accelerators or sensory front-ends, they can continuously compare transient runtime information to identify temporal correlations, recurring patterns, or previously observed states. Visual place recognition (VPR) is a representative example, where feature matching corresponds to nearest-neighbor search in Hamming space and, together with feature extraction, can account for approximately 60--70\% of execution time \cite{vemulapati2022orbbasedslamacceleratorsoc}. Similar requirements arise in transformer attention and token reuse mechanisms \cite{manea2025gaincellcam,molomochir2025camformerassociativememoryneed}, \ac{CAM}-based learning hardware with continuously updated weights or prototypes (Sec.~\ref{sec:training}), and the second stage of two-stage genomic search, where candidate locations returned by a Shift\ac{CAM} filter differ for every query and, therefore, must be written before each fine-grained alignment (Sec.~\ref{sec:app-specificCAM}). The Shift\ac{CAM} filter that stores the reference database itself belongs to the static tier below. Despite these broad and highly-relevant applications, dedicated dynamic \ac{CAM} architectures supporting rapidly updated representations remain comparatively under-explored.
    
    \paragraph{Tier-2 Static / Search-intensive \acp{CAM}}
    Static \acp{CAM} are large-capacity associative memories optimized for denser storage and energy-efficient search. In contrast to dynamic \acp{CAM}, write performance is not as critical as the applications have infrequent programming operations, but  prioritize search energy, storage density, and scalability. They are therefore particularly well suited for comparatively static datasets that are queried frequently, but only occasionally updated. This application class is currently the most established domain for \ac{CAM}-based acceleration, with numerous reported demonstrations across both classical and emerging workloads. Classical examples include routing and packet-classification tables, where \acp{CAM} have long been used for fast lookup operations \cite{mcauley1993fast,zane2003coolcams}. Static search memories have also been explored for database search and large-scale nearest-neighbor retrieval \cite{Lee1993InexactCAM,34085}. More recent application directions include tree-based search accelerators, discussed in Sec.~\ref{sec:CAM_trees}, and genomics workloads, discussed in Sec.~\ref{sec:genomics} \cite{Pedretti2021Tree,laguna2020seed,he2024shiftcam}.

    \subsection{Scaling \acp{CAM} Arrays with New Materials and Vertical Integration}

    Scaling \acp{CAM} to larger capacities exposes a central device--circuit trade-off between sensing margin and storage density. Robust \acp{CAM} require clear match--mismatch discrimination to tolerate device variability, noise, and non-ideal search dynamics, but larger sensing margins typically demand additional transistors, stronger isolation, or peripheral gain stages, all of which increase cell area. Highly compact cells instead operate with smaller margins and fewer states per cell, making them more sensitive to variability and array-level parasitics \cite{Bazzi2022,Manea2026SALM}. The trade-off becomes more severe in large arrays, where sensing margin, interconnect loading, match-line capacitance, and search bandwidth turn into coupled system-level constraints rather than isolated cell-level metrics. Relaxing this trade-off requires devices and integration schemes beyond planar silicon, and two-dimensional materials are the clearest example at the device level.

    The appeal of 2D semiconductors starts at the device level. Their atomically thin body improves electrostatic control, suppresses short-channel leakage, and reduces parasitic capacitance, which sharpens the electrical contrast between matching and mismatching states during search while limiting unwanted current paths \cite{lin2023recent,liu2021promises,lu2024projected}. The most direct benefit for \acp{CAM} is the drain capacitance that every cell attaches to the match line and that is charged and discharged on every search: a thin body on an insulator has no bulk junction to deplete, lowering both the $CV^2$ search energy and the match-line RC delay, although the delay benefit requires the read current to be maintained. Recent ultrafast 2D flash studies show that this read current depends more strongly on contacts and interfaces than on the channel material itself, and that the bottleneck can be relaxed through semimetal and edge-contact strategies that reduce contact resistance and weaken Fermi-level pinning \cite{liu2021ultrafast,wu2021atomically,yu2023simultaneously,das2013high,liu2022fermi,li2023approaching,shen2021ultralow}. Semimetal-contacted MoS\textsubscript{2} flash a\acp{CAM} illustrate the direction, combining high read current with a large ON/OFF ratio for low-latency analog search \cite{gao2025sb}.

    These device properties translate into a system-level opportunity, since back-end-of-line (BEOL) compatibility, shared by 2D channels and by thin-film oxide semiconductors, permits stacking memory and search layers directly above CMOS logic. Vertical integration shortens the wiring between storage and peripherals and thereby reduces interconnect parasitics, but it does not eliminate the underlying compromises between density, retention, programmability, and sensing robustness. It is therefore most valuable wherever capacity, bandwidth, and energy are limited by footprint and wiring rather than by the memory cell alone \cite{qiu2025monolithic,liu2021promises,lu2024projected}.

    Both tiers of the associative hierarchy introduced in Sec.~\ref{sec:memory-hierarchy} can benefit from this, but they call for different stacked platforms. For the static tier, the 2D flash devices discussed above are the natural candidate, since retention and density dominate and their dual-gate structure already suggests vertically stacked complementary cells that form a compact a\ac{CAM} without an external analog inverter \cite{gao2025sb}. The dynamic tier instead requires fast and endurance-tolerant writes, which points to the gain-cell a\acp{CAM} of Sec.~\ref{sec:tech-landscape} implemented in oxide-semiconductor platforms such as IGZO. Here the low off-current of the oxide access transistor extends the retention time of the dynamically stored key and thereby relaxes the refresh overhead, while BEOL-compatible processing allows multiple such layers to be stacked above a shared CMOS search periphery, reducing the effective area per stored bit relative to a planar cell. This combination could therefore enable monolithic 3D integration of dense vertical dynamic \ac{CAM} arrays \cite{9401798,10268258,lu2024high,shi_counteractive_2023,10185398,ye_double-gate_2020}.

    The open question is therefore not whether these materials can demonstrate \ac{CAM} behavior, but whether they can be manufactured as scalable 3D associative systems. This requires stable analog programmability, a controllable memory window, adequate retention and endurance, and low device-to-device variation so that stored match ranges remain predictable across large arrays \cite{liu2021ultrafast,yu2023simultaneously,jiang2024scalable,gao2025sb}. Equally important are wafer-scale synthesis, layer-to-layer alignment, thermal-budget management, and the co-integration of selectors, sense amplifiers, and digital control \cite{jiang2024scalable,li2021epitaxial,liu2021promises}. If these hurdles can be reduced, new materials may matter less as standalone device novelties than as enabling platforms for dense monolithic 3D \ac{CAM} systems.

    \section{Applications}
    Associative retrieval is a fundamental operation that appears across many computational workloads. In this section, we survey how \acp{CAM} can accelerate applications ranging from thresholding and tree-based inference to genomic sequence alignment and transformer attention. We further discuss emerging directions in trainable and differentiable \acp{CAM}, highlighting both the opportunities and challenges of learning directly within associative memories. Together, these examples illustrate the growing role of CAMs as a general-purpose substrate for associative computation.
    
    \subsection{Tree Model Inference with CAMs}
    \label{sec:CAM_trees}

\begin{figure}
        \centering
        \includegraphics[width=1\textwidth]{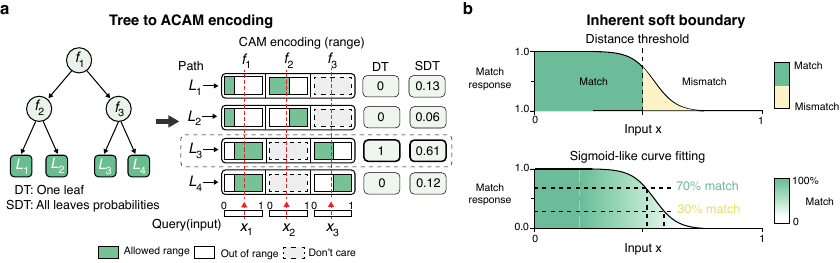}
        \caption{\textbf{Mapping decision-tree inference onto analog \ac{CAM} hardware.} (a) A tree-based model can be mapped to \ac{CAM} by a range encoding way. Each path of the tree model corresponds to a match-line row in the CAM array. The ``less than" or ``greater than" in tree nodes maps to the range stored in \ac{CAM} cells. Traditional Decision Trees (DT) make decisions only one leaf at a time. However, a Soft Decision Tree makes decisions through all leave probabilities across all paths, selecting the highest one. (b) The inherent soft boundary for a range stored in \ac{CAM} is usually simplified by a defined distance threshold to determine the match or mismatch results. However, we observed that the inherent soft boundary is sigmoid-like and can be well fitted by sigmoid functions. We can exploit its property to obtain continuous match degrees like 30\% or 70\% rather than binary match or mismatch. This probability can also be used in Soft Decision Tree nodes as gating functions.}
        \label{fig:tree_to_CAM}
    \end{figure}

    Beyond database-style lookups, inference of tree models (Fig.~\ref{fig:tree_to_CAM}) is one of the clearest examples of a workload mapping naturally onto \ac{CAM} hardware. Decision trees, random forests, and related rule-based learners are composed of repeated inequality tests and threshold comparisons rather than dense linear algebra. However, their inference remains inefficient on conventional processors because tree traversal induces irregular memory accesses, branch divergence, and sequential data dependencies. \ac{CAM} arrays provide a more direct substrate for these workloads: node thresholds, path conditions, or leaf regions can be encoded in memory and evaluated in parallel, exposing comparison-level parallelism that is largely hidden in software-based traversal \cite{Pedretti2021Tree,10753423,yin2024deep,rakka2023dt2cam,pedretti2022general}.

    Tree inference is especially attractive because it provides a direct path from conventional machine-learning training to \ac{CAM} deployment. Unlike approaches that require differentiable memory access during end-to-end optimization, tree models can be trained with standard algorithms and only mapped to hardware after training. In a tree model, each internal node tests whether a specific input feature falls above or below a learned threshold, and each root-to-leaf path defines a decision rule as the conjunction of these inequalities, as illustrated in Fig.~\ref{fig:tree_to_CAM}(a). The learned inequalities or decision boundaries can then be translated almost one-to-one into stored \ac{CAM} ranges \cite{Pedretti2021Tree,10753423}. This makes \ac{CAM}-based tree inference compatible with conventional offline training, while retaining the interpretability of rule-based models and shifting the main hardware challenge to accurate programming and verification of the search boundaries to mitigate non-idealities \cite{manea2023non, 10848806}

    A particularly promising direction is to move beyond hard trees. Recent work maps soft decision trees to analog \ac{CAM} hardware by exploiting the intrinsic soft boundaries of threshold-tunable devices, turning a device limitation into a probabilistic compute primitive that improves robustness to variation and perturbation \cite{wen2026trustworthy}. More broadly, \ac{CAM}-based tree inference should co-design model classes with analog device physics rather than forcing analog comparisons to emulate abrupt digital thresholds.
    
    \subsection{DNA Genomics}
    \label{sec:genomics}
    
    \begin{figure}[htb!]
    \centering
    \def\svgwidth{1\textwidth}
    \fontsize{8}{9.6}\selectfont
    \includegraphics[width=1\textwidth]{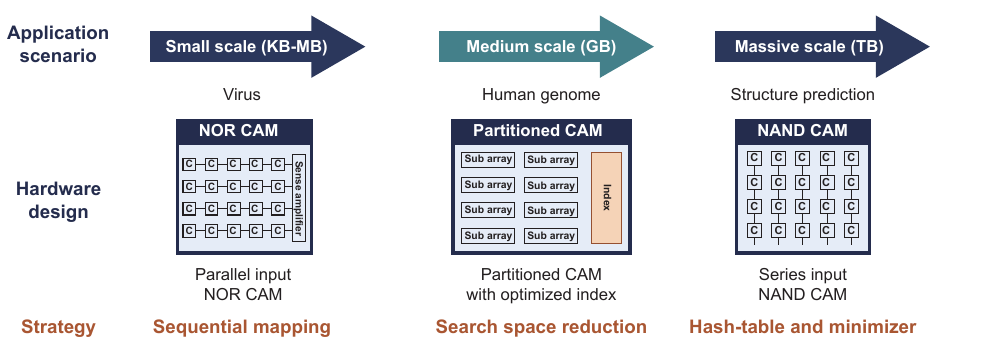}
    \caption{\textbf{\ac{CAM} architecture selection across genomic database scales.} The choice of \ac{CAM} design depends on the genomic database. For a small virus reference genome, a fast NOR \ac{CAM} is preferred. For the human genome (around 3 GB), some form of search space reduction is essential. For searching extremely large databases, a NAND-based \ac{CAM} with software–hardware co-design like hash tables is preferred.}
    \label{fig:genomics}
    \end{figure}

    Current DNA sequencing machines only extract a small part of a DNA sequence from the complete genome. Thus, the first step is to map these short reads to a large reference. To overcome the bottleneck of data movement, \ac{CAM} has been widely used in bioinformatics applications such as virus detection\cite{hanhan2022edam, zhong2023asmcap, jahshan2023dash, he2024shiftcam}, metagenomics\cite{ghiasi2024megis}, and real-time analysis\cite{he2025real}. However, different species show different genetic complexity, biological databases vary significantly in size, showing that hardware acceleration strategies must be tailored to specific data scales. For small-scale datasets, such as viral databases ranging from kilobytes (KB) to megabytes (MB), high-speed NOR \ac{CAM} architectures are highly effective due to their rapid, parallel search capabilities\cite{hanhan2022edam, zhong2023asmcap, jahshan2023dash, he2024shiftcam, jiang2026dshd}, and they usually map DNA sequences sequentially into the \ac{CAM} for best performance, but at the cost of substantial memory redundancy. However, as the data scale expands to gigabyte-level (GB) human genomic databases\cite{laguna2020seed, liu2023reconfigurable}, area and energy constraints make sequential storing and direct searching impractical. Consequently, researchers have implemented optimized indexing and \ac{CAM} partitioning schemes to reduce the active search space, effectively balancing storage capacity and search efficiency\cite{morris2026np}. Furthermore, emerging fields like protein structure prediction and metagenomics typically require datasets exceeding terabytes (TB). This extreme scale makes conventional NOR architectures impractical, requiring high-density NAND architectures with hardware-software co-design techniques, such as hash tables and minimizers\cite{ghiasi2024megis}.

    Beyond storage capacity difference, different applications also require different balance between analysis speed and accuracy. Considering genetic mutations and sequencing errors in sequencing data, approximate matching in \ac{CAM} is essential, which is typically achieved by tuning the reference voltage of sense amplifiers\cite{hanhan2022edam, zhong2023asmcap, jahshan2023dash, he2024shiftcam, he2025real}. And for Next-Generation Sequencing (NGS) data with low error rates ($\sim$0.01\%), designs like EDAM\cite{hanhan2022edam} and Asmcap\cite{zhong2023asmcap} use complicated combinational logic circuit in \ac{CAM} cell to tolerate single insertion or deletion (indel) for better performance. However, Third-Generation Sequencing (TGS) technologies typically have much higher error rates (up to 5\%), designs such as ShiftCAM\cite{he2024shiftcam, jiang2026dshd} can tolerate multiple indels by implementing shift logic in time-domain. Furthermore, mapping Levenshtein automata directly into T\ac{CAM} arrays\cite{graves2020memory} can be a highly accurate approach to compute edit distances for approximate matching, but this process needs some iterations and would sacrifice speed efficiency. For targeted sequencing, real-time control makes speed the top priority. To achieve this, researchers proposed an LSH-T\ac{CAM} system to extract binary features from nanopore raw signals and search them in \ac{CAM} directly instead of conventional base pair\cite{he2025real}. Although this is the fastest method, it sacrifices some accuracy.

    \subsection{Transformer Attention with \acp{CAM}}
    \label{sec:CAM_attetion}
    
    Today's large language models (LLM) depend on Transformer attention \cite{vaswani2023attentionneed}, which can be understood as a form of associative memory.  This has conceptual roots in Kohonen's correlation matrix memory \cite{Kohonen1972}, where key--datum pairs are stored as outer products and recalled by multiplying the resulting memory matrix with a retrieval key. In modern terminology, this resembles an unnormalized key--value retrieval operation. This connection has been further developed in modern Hopfield networks, which interpret the query--key interaction in transformers as associative retrieval over stored patterns \cite{ramsauer2021hopfieldnetworksneed}. More broadly, several works have shown that neural networks and attention layers can be viewed as complementary forms of associative memory, linking representation learning, pattern retrieval, and transformer computation within a common conceptual framework \cite{radhakrishnan2020overparameterized,zhong2025understandingtransformerperspectiveassociative,bietti2023birth}.
    
    While transformer attention relies almost exclusively on scaled dot-product similarity, recent studies suggest that alternative similarity measures--such as Manhattan and Euclidean distance--can offer improved retrieval robustness and practical memory capacity in associative memory systems \cite{Millidge2022UniversalHopfield}. The dominance of scaled dot-product attention may therefore partly reflect its efficient mapping onto GPU-based linear algebra, rather than an inherent optimality for associative retrieval. This observation motivates the exploration of \ac{CAM}-based architectures, which naturally support distance-based matching and can implement highly selective similarity metrics directly in memory.

    Recent works have demonstrated that transformer attention mechanisms can be implemented efficiently using \ac{CAM}-based architectures \cite{manea2025gaincellcam, molomochir2025camformerassociativememoryneed}. In these approaches, the generation of attention scores, corresponding to the query-key (QK) lookup operation, is implemented directly within a \ac{CAM} structure. This enables content-based similarity search to be performed  in-memory, avoiding the explicit computation of large matrix multiplications and softmax operations typically required in conventional dot-product attention, thereby significantly simplifying the underlying compute-in-memory hardware architecture.

    Existing pre-trained transformer models cannot be directly transferred to \ac{CAM}-based attention architectures, but instead require some retraining. A central challenge is that gradient-based optimization requires differentiable memory access and backward signal propagation whereas most \ac{CAM} operations are inherently non-differentiable and provide no natural backward path. Existing approaches therefore rely on surrogate models or approximate gradients, as discussed in Section~\ref{sec:training}, but these methods can reduce training stability, scalability, and accuracy compared with standard dot-product attention \cite{graves2016differentiable,vaswani2023attentionneed,manea2025gaincellcam}. A promising alternative is non-differentiable retrieval-based attention, where associative lookup is decoupled from gradient propagation, as in k-NN-based Memorizing Transformers and recent sparse-attention methods that selectively retrieve relevant KV-cache entries \cite{wu2022memorizingtransformers,deepseekai2025deepseekv32pushingfrontieropen,yuan2025nativesparseattentionhardwarealigned}. This paradigm is particularly well suited to \ac{CAM}-based hardware.
    
    Another challenge is that \ac{CAM}-based search can become increasingly energy-intensive as the KV cache grows with the transformer context length. Sparse attention reduces this cost by evaluating only selected tokens, while \acp{CAM} can act as a preselector or sorting engine to identify relevant KV entries before subsequent computation \cite{child2019generatinglongsequencessparse,roy2021efficient,Hybrid_CAM_SRAM_sparse_attention}. However, these approaches typically retain a flat, globally accessible search space. Hierarchical associative structures could instead progressively route each query toward increasingly selective memory subsets, thereby reducing unnecessary array activation and improving scalability. The corresponding \ac{CAM} architectures and energy–latency trade-offs are discussed in Section~\ref{sec:tree}.
    
    \subsection{Machine Learning with \acp{CAM}}
    \label{sec:training}

    \begin{figure}[htb!]
        \centering
        \includegraphics[width=1\linewidth]{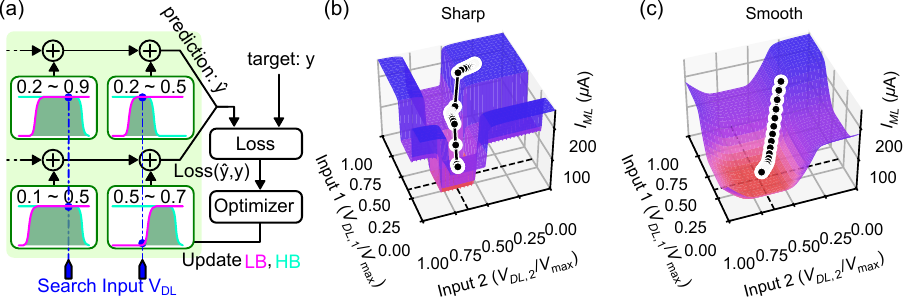}
         \caption{\textbf{Machine learning with analog \acp{CAM}.}
        (a) Surrogate-model training pipeline for learning analog \ac{CAM} parameters using TorchCAM, where the \ac{CAM} decision boundaries are optimized through standard backpropagation.
        (b) Comparison between \acp{CAM} with sharp decision boundaries, which provide high selectivity but limited adaptability, and smoother response functions that improve differentiability and trainability.
        (c) Representative approaches for mapping learned parameters onto \ac{CAM} hardware.}
        \label{fig:CAM_learning}
    \end{figure}
    
    Many \ac{CAM} applications discussed in this perspective require training methodologies compatible with gradient-based optimization. Depending on the deployment scenario, this may involve differentiable software models of the \ac{CAM} hardware for offline training, or training and fine-tuning directly on the hardware itself. In such systems, the memory operation itself must often be differentiable such that gradients can propagate through the associative memory during backpropagation \cite{graves2016differentiable}. The required degree of differentiability depends on the application. Tree models can be trained using conventional algorithms and subsequently mapped directly onto \ac{aCAM} ranges without backpropagating through a hardware model, as discussed in Section~\ref{sec:CAM_trees} \cite{Pedretti2021Tree,10753423}. Hardware non-idealities can then be mitigated during deployment through decision-boundary-aware programming and verification. Because many device- and CMOS-level process variations in \ac{aCAM} are static, their effects can be absorbed during programming by calibrating each memory cell to the desired inequality boundary rather than targeting identical physical device states \cite{manea2023non,10848806,11563370}.

    Some applications fundamentally require differentiable associative retrieval. Transformer attention, as outlined in Section~\ref{sec:CAM_attetion}, is an important example \cite{vaswani2023attentionneed,graves2016differentiable}. Here, an accurate and computationally efficient model of the \ac{CAM} hardware is essential. One approach is to approximate the inherently sharp decision boundaries with smooth functions, as implemented in TorchCAM, where shallow sigmoid functions model the boundaries and are progressively sharpened during training, thereby providing surrogate gradients for backpropagation, as shown in Fig.~\ref{fig:CAM_learning}(a) \cite{Manea2022TorchCAM}. Similarly, the L2S algorithm presented in \cite{Pedretti2022differentiableCAM} optimizes memristor conductances using gradients computed from a differentiable surrogate circuit model, enabling compensation of hardware non-idealities such as match-line crosstalk during training \cite{Manea2026SALM}. However, these approaches still restrict learning and backpropagation to software-based models rather than the physical hardware itself. If device-specific variations and non-idealities are to be compensated directly, actual in-hardware learning or fine-tuning becomes necessary.

    Training directly on \ac{CAM} hardware remains a major challenge because most \ac{CAM} architectures provide little or no usable gradient information for backpropagation. Differentiable \acp{CAM} offer a potential route to overcoming this limitation by employing smooth, shallow decision boundaries that support gradient-based optimization \cite{Pedretti2022differentiableCAM}. However, this increased differentiability typically comes at the cost of reduced storage capacity and selectivity \cite{Manea2026SALM}. Figure~\ref{fig:CAM_learning}(b) compares sharp and smooth \ac{CAM} decision landscapes, highlighting the improved gradient propagation enabled by the smoother landscape. Another major challenge is that all \ac{CAM} hardware does not provide a natural backward path for gradient propagation. Enabling in-hardware backpropagation therefore requires substantial circuit modifications at the \ac{CAM} cell level to support transposed \ac{CAM} operations, dramatically increasing the area footprint. One possible alternative to full in-hardware backpropagation is to employ local or forward-only learning rules, such as Hebbian-style updates \cite{lansner2025benchmarkinghebbianlearningrules}, frequency-multiplexed gradient descent \cite{Multiplexed_gradient, Boon_gad_desc_homodyne2025}, or node-perturbation-based training \cite{dalm2026nodeperturbationeffectivelytrain}. By reducing or eliminating the need for dedicated backward-propagation paths, these approaches may be particularly attractive for \ac{CAM}-based hardware. Although such approaches have received little attention in the context of \acp{CAM}, they may provide a more hardware-compatible learning paradigm for future \ac{CAM}-based architectures.

    From a broader perspective, several challenges must be addressed to enable \ac{CAM}-based \ac{AI} systems. First, accurate and computationally efficient software models are required to support standard gradient-based optimization and backpropagation. Second, specialized \ac{CAM} circuits capable of supporting in-hardware learning and backward signal propagation will likely be necessary, although such approaches are expected to incur additional area overhead and require dynamically programmable memory structures. Finally, alternative local learning rules should be further explored to compensate hardware non-idealities and alleviate the need for full global backpropagation in \ac{CAM}-based architectures.

\section{From \ac{CAM} Arrays to Associative Computing Architectures}

Individual \ac{CAM} arrays provide the fundamental operation of content-based search, but practical systems require scalable architectures. This includes architectural decisions on where \acp{CAM} are placed, how they interact with surrounding compute and memory blocks, how data movement is organized, and which memory technologies are assigned to different levels of the memory hierarchy. At the same time, the increasing diversity of associative hardware requires correspondingly flexible software tools capable of designing, evaluating and optimizing these heterogeneous systems. Together, these architectural and software considerations connect the device, circuit, and application perspectives introduced above. They also demonstrate that \acp{CAM} are not restricted to a single integration strategy, but can serve as standalone retrieval engines, associative memory tiers, near-memory accelerators, or tightly integrated components within broader \ac{CIM} systems. In the next section, we discuss how \acp{CAM} can be designed, integrated and scaled as building blocks of future associative computing architectures.

    \subsection{Design Automation for Large-Scale Associative Computing Architectures}
    \paul{added this subsection as i think it is important also good way to citer Sharon Hus evaCAM}
    The growing diversity of \ac{CAM} technologies, search modalities, and system architectures creates an increasingly large design space spanning devices, circuits, architectures, and applications. Exploring this space manually is becoming impractical, particularly as different memory technologies exhibit distinct latency, energy, area, precision, and endurance trade-offs. Consequently, future associative computing systems will require mature design automation frameworks capable of circuit- and architecture-level evaluation, rapid design-space exploration, hardware-aware mapping, and compiler support.
    
    Recent efforts already point in this direction. Eva-CAM~\cite{11150515} enables technology-aware evaluation of heterogeneous \ac{CAM} architectures across multiple abstraction levels, facilitating rapid exploration of architectural trade-offs. At the application level, X-TIME~\cite{10753423} demonstrates hardware-aware mapping and optimization of tree-based machine-learning models onto analog \ac{CAM} architectures, while compiler frameworks such as C4CAM~\cite{Farzaneh2024} further generalize this concept by translating high-level programs into optimized implementations for heterogeneous \ac{CAM} accelerators. Together, these efforts represent important first steps toward a complete hardware--software ecosystem for associative computing, analogous to the role that compiler and programming frameworks have played in enabling modern GPU computing.
    
    \subsection{Application Specific CAM Architectures}
    \label{sec:app-specificCAM}
    Many applications require broader operations than matching between queries and stored \ac{CAM} entries. Depending on the workload, associative search may need to support approximate similarity, shifted alignment, programmable precision, winner-take-all selection, or direct retrieval of associated values. These functions require application-specific peripheral circuitry and architectural organization beyond the \ac{CAM} array itself. The following examples illustrate how such specialization extends basic lookup into more capable associative computing systems.
    
    \paragraph*{ShiftCAM for Genomic Matching}
    Unlike conventional similarity search, there are frequent insertion and deletion (``indel") errors in sequencing data, which will decrease conventional \ac{CAM} performance. This is because conventional \ac{CAM}s search according to Hamming distance. However, even a single indel error can lead to a large Hamming distance, making it fail to represent sequence similarity. As discussed in Sec.~\ref{sec:genomics}, Shift\ac{CAM} uses additional time-domain shift logic circuits to tolerate indel errors, which is efficient but not accurate enough for many practical applications. To further improve search accuracy, additional architectural support is required. A promising direction is to use Shift\ac{CAM} as a first-stage filter to quickly identify rough candidate locations in the reference database, and then use a T\ac{CAM}-based second stage to perform fine-grained edit distance matching through Levenshtein automata. This second stage can evaluate sequence similarity more accurately, but the candidate locations generated by Shift\ac{CAM} differ across input queries, thus additional dynamic memory is needed to store and update the input-dependent candidate DNA fragments and automaton states for Levenshtein automata. By implementing this dynamic memory with BEOL compatible materials such as IGZO or 2D materials\cite{qiu2025monolithic}, this potential architecture could be realized through high density 3D integration, further improving efficiency.
    
    \paragraph*{Soft CAM Architecture}
    A second application-specific direction is to use \acp{CAM} for probabilistic inference, where preserving graded match information is more important than collapsing every comparison into a binary decision. As discussed in Sec.~\ref{sec:CAM_trees}, soft decision tree models are a natural example because inference operates by accumulating probabilities across multiple candidate paths rather than selecting a single hard branch \cite{wen2026trustworthy}. In these workloads, application-specific architectures are needed because conventional \ac{CAM} optimized for exact or thresholded matching does not preserve the analog information required by the model. A Soft CAM therefore maps candidate paths to rows, uses the cell responses to evaluate soft node conditions, and accumulates the resulting analog signals on the match line so that each row carries a probability-like confidence score. After calibrated sensing, the most likely outcome can be selected through winner-take-all readout. Since this winner-take-all operation can be implemented directly in hardware as part of the \ac{CAM} architecture, it can avoid additional digital post-processing and improve overall efficiency \cite{dube2025analog}. A key architectural requirement is to preserve the relative ordering of row-level probabilities, even under device variations and finite-slope readout, rather than only maximizing binary match discrimination. This Soft CAM example illustrates the benefits for future associative systems to be tailored for workload-specific organization, beyond justbasic \ac{CAM} primitives.
    
    \subsection{Integrating \acp{CAM} into \ac{CIM} Architectures}
    
    \Acp{CAM} can be viewed as the missing associative primitive in future compute-in-memory (\ac{CIM}) systems. Crossbar arrays efficiently accelerate linear algebra, but they do not natively provide content-based retrieval. \acp{CAM} fill this gap, enabling associative recall directly in memory. This is particularly relevant because many analog \ac{CIM} architectures are dominated by data-conversion overhead, with \acp{ADC} and \ac{DAC} reported to contribute roughly 60--90\% of the energy consumption and up to $\sim$80\% of the area \cite{shafiee2016isaac,Ankit2019PUMA,yao2020fully}. By accepting analog inputs and producing digital match outputs, analog \acp{CAM} could reduce repeated conversions while adding search as a native compute operation. We next discuss two architectures that move toward this tighter integration of associative search and crossbar-based \ac{CIM}.
    
    \begin{figure}[htb!]
        \centering
        \includegraphics[width=1\linewidth]{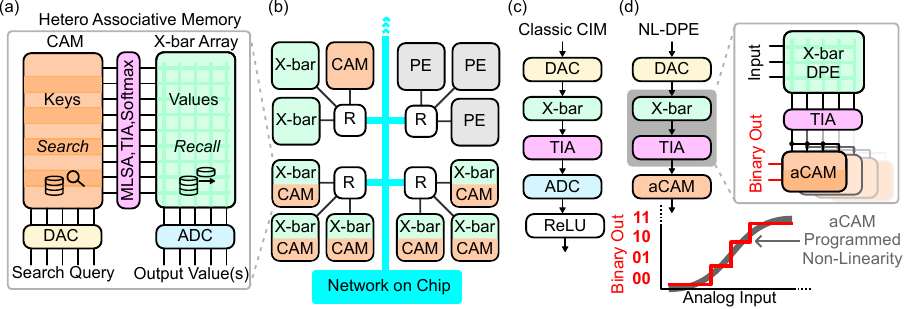}
        \caption{\textbf{Integrating \acp{CAM} into \ac{CIM} architectures.} (a) Hetero-associative memory, where a \ac{CAM} stores the keys and a crossbar array stores the values, jointly implementing a lookup table analogous to an Attention Mechanism.
        (b) Example system-level architecture with distinct \ac{CAM} and \ac{CIM} processing elements (PEs), as well as hybrid hetero-associative / attention units connected with a network on chip.
        (c) Conventional \ac{CIM} inference flow, where digital inputs are converted to analog signals using \acp{DAC}, computations are performed in-memory, and analog outputs are digitized by \acp{ADC}.
        (d) Non-linear dot-product engine (NL-DPE) flow, where a\acp{CAM} replace ADCs. Architecture of the NL-DPE, illustrating binarization of crossbar outputs and the embedding of non-linear activation functions within the CAM. Instead of fully digitizing $y=f(x)$, the analog signal is compared against stored ranges which are memory-resident and programmable, yielding a digitized output signal.}
        \label{fig:cam_rchtiectures_scale_up}
    \end{figure}
    
    \paragraph{Hetero-associative Memories with \acp{CAM} and Crossbars}
    
    The combination of \ac{CAM} and \ac{CIM}, shown in Fig.~\ref{fig:cam_rchtiectures_scale_up}(a), realizes a hetero-associative memory by mapping input keys to stored values. The \ac{CAM} performs associative key matching, while the \ac{CIM} crossbar (X-bar) array stores and retrieves the corresponding values, together implementing a hardware \ac{LUT}. By tightly integrating both arrays, the analog output of the \ac{MLSA} or \ac{TIA} can directly drive the \ac{CIM} array, eliminating intermediate \acp{ADC} and preserving an end-to-end analog signal path. Given the wide range of existing implementations and its conceptual generality, this structure emerges as a fundamental building block for future \ac{CAM}-enhanced \ac{CIM} architectures.
    
    Applications of this principle have already emerged across multiple domains. In finite-state machines, \acp{CAM} detect the current state while \ac{CIM} arrays encode the corresponding state-transition \cite{Graves2020FSM}. Another prominent example is the Attention Mechanism, which can itself be interpreted as a hetero-associative unit linking queries to values through coupled \ac{CAM} and \ac{CIM} arrays \cite{manea2025gaincellcam,molomochir2025camformerassociativememoryneed}, discussed further in Sec.~\ref{sec:CAM_attetion}. For example, in decision tree (tree model) implementations, the \ac{CAM} evaluates decision rules performing a classification, while a crossbar array retrieves the class label associated with the selected leaf node \cite{Pedretti2021Tree}. A  recent work applied this concept to Kolmogorov-Arnold Networks (KANs), where \ac{CAM}-based LUTs are used to encode nonlinear activation functions within the network \cite{Yu_CAM_LUT_KA}. Extending this concept further, networks of hetero-associative units or \acp{LUT} can, in principle, realize arbitrary computations. Modern FPGAs represent the closest practical realization of this paradigm, employing large arrays of configurable LUTs and programmable interconnects to implement digital logic rather than relying solely on fixed-function processors \cite{Kuon_FPGA_2008}.

    A potential system-level realization is shown in Fig.~\ref{fig:cam_rchtiectures_scale_up}(b), where heterogeneous \acp{PE}, including conventional crossbar-based \ac{CIM} units, standalone \acp{CAM}, and hybrid hetero-associative \ac{CAM}–\ac{CIM} units, are integrated on-chip and interconnected through a network-on-chip.
    
    \paragraph{Non-Linear DPE for ADC-Light Crossbar Readout}
    In classical mixed-signal \ac{CIM} accelerators, as shown in Fig.~\ref{fig:cam_rchtiectures_scale_up}(c), the bottleneck is not only the matrix-vector multiplication itself, but also the analog-to-digital conversion of the accumulated output \cite{shafiee2016isaac,Ankit2019PUMA,yao2020fully} and the subsequent nonlinear mapping in the digital domain. A second integration pattern for \acp{CAM} and crossbar arrays therefore operates in the reverse direction: the crossbar array first performs the analog dot product, while the \ac{CAM} replaces the conventional \ac{ADC} and digital post-processing pipeline. This forms a non-linear dot-product engine (NL-DPE) \cite{zhao2025nldpeanaloginmemorynonlinear,Zhao2023RACEITAR,zhao2024noise}, as illustrated in Fig.~\ref{fig:cam_rchtiectures_scale_up}(d).

    \acp{CAM} provide a natural way to embed this readout stage in memory. Instead of fully digitizing the crossbar output, the analog signal can be compared directly against stored thresholds or intervals. The \ac{CAM} then produces a binary code indicating the region of the programmed nonlinearity in which the input signal falls, which can be passed to subsequent processing stages. This architecture is conceptually related to a flash \ac{ADC}, but replaces fixed comparator levels with programmable, memory-resident thresholds that can be adapted to the target nonlinear activation function, the signal statistics, or even the calibration of the \ac{DPE} \cite{Hong2025AdaptiveADC,jung2026hybrid}. From this perspective, NL-DPE can merge analog-to-digital conversion and nonlinear activation into a tighter analog front end, potentially reducing \ac{ADC} overhead while enabling application-specific post-processing directly in memory. Whether this approach is beneficial depends on the required resolution, the number of stored ranges, sensing cost, variation tolerance, and whether associative thresholding scales more favorably than a conventional \ac{ADC} followed by digital logic \cite{Li2020,Hong2025AdaptiveADC,zhao2025nldpeanaloginmemorynonlinear}.

    \subsection{Reducing Search Energy Through Hierarchical \ac{CAM} Architectures}
    \label{sec:tree}

    Recent work has shown that large-scale associative retrieval can benefit substantially from hierarchical memory organization.  Flat retrieval architectures scale poorly with increasing memory size, suggesting that future (larger) memory systems will require structured memory navigation mechanisms beyond dense global search, further evidenced in the example of retrieval augmented generation (RAG) \cite{rezazadeh2025isolatedconversationshierarchicalschemas}. This can also be attractive for the implementation of KV cache within the attention mechanism of transformers (Section~\ref{sec:CAM_attetion}), which has been demonstrated theoretically in \cite{chalkidis2022explorationhierarchicalattentiontransformers}. As introduced in Box~\ref{box:hist_brain}, sparse hierarchical retrieval may explain how the brain accesses relevant associations within vast stored memory spaces without requiring exhaustive and energy-hungry search. Tree-based inference, discussed in Section~\ref{sec:CAM_trees}, is typically implemented using parallel associative search. However, the underlying tree structure can also be exploited to enable hierarchical and sparse search, reducing the number of candidate paths that must be evaluated.

    Early CAM works identified the large energy consumption associated with fully parallel search operations where all highly capacitive search-lines and match-lines switch during every query \cite{1327749}. To mitigate this problem, segmented and hierarchical \ac{CAM} architectures were proposed, where search operations can terminate early once any mismatch is detected, reducing unnecessary switching activity and lowering overall power consumption by up to 60\% \cite{1327749}. This principle of progressive pruning has been extended directly to search-lines, where data-aware gating nodes actively halt signal propagation upon encountering "don't care" bits in IP routing tables to cut energy consumption by nearly half \cite{mathan2016data}. Similar ideas were recently validated in application-driven associative computing, where sequential feature-wise search with early termination reduced Random Forest inference energy by approximately 50\% across multiple datasets while requiring only three to four sequential feature evaluations per search in most cases \cite{Manea2026SALM}. While a\acp{CAM} enable highly parallel Random Forest inference, sequentially evaluating decision paths from root to leaf allows mismatching candidates to be discarded early, substantially reducing switching activity and search energy. 

    More generally, the above results highlight that pipe-lined or sequential associative search strategies can effectively trade-off search latency for energy efficiency. This may become particularly important for transformer attention mechanisms, where associative search operations potentially scale with the entire context window and therefore incur rapidly increasing search energy and bandwidth requirements. Nevertheless, hierarchical associative search from a hardware-centric \ac{CAM} perspective remains largely unexplored, despite multiple works already demonstrating the strong potential of hierarchical and staged \ac{CAM} architectures for reducing search energy and improving scalability.

    We advocate for hierarchical search to be treated as a core architectural design target in future \ac{CAM} hardware. Rather than activating the full array for every query, future systems should support sparse search directly in hardware, for example through row-wise or block-wise activation schemes in which only candidate regions are enabled after an earlier search stage reports a match.

    \section{Conclusions and Future Directions}
       
    In this perspective, we propose \acp{CAM} as a foundational hardware primitive for advanced \ac{AI} systems and broader applications, not just as specialized lookup circuits in networking or database systems. We showcase recent advancements across emerging memory technologies, algorithmic workloads, and novel architectures for broad application spaces. We have examined this paradigm shift across many levels: underlying device technologies, which determine density, robustness, and sensing limits; the application perspective, which highlights emerging CAM workloads such as attention, tree-based inference, and genomics; and the architectural perspective, which positions \acp{CAM} as a natural complement to modern \ac{AI} computing infrastructure.

    \begin{figure}[htb!]
        \centering
        \includegraphics[width=1\linewidth]{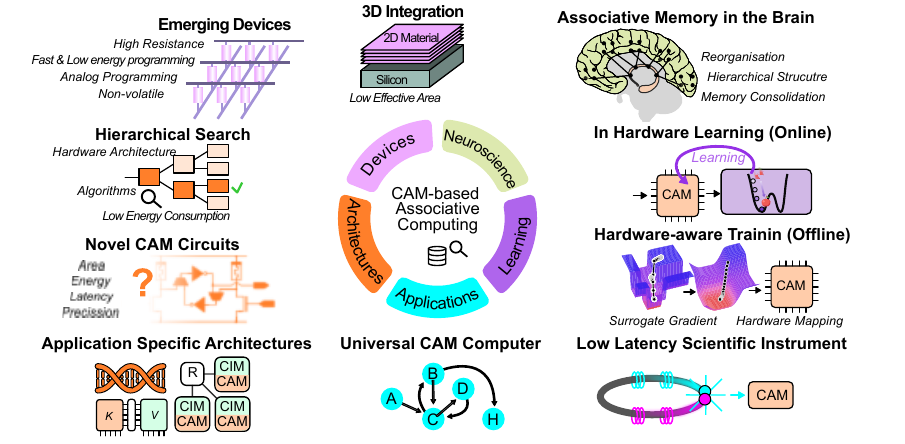}
        \caption{\textbf{Roadmap toward scalable \ac{CAM}-based associative computing.}
        Realizing associative computing as a general-purpose computing paradigm requires progress across multiple levels of the technology stack, from emerging memory devices, circuits, and architectures to learning methods, neuroscience-inspired memory organization, and application-driven system design. The figure highlights key research opportunities, including hierarchical search, monolithic 3D integration, hardware-aware and online learning, application-specific \ac{CAM} architectures, and emerging domains such as probabilistic computing and low-latency scientific instrumentation.}
        \label{fig:core_future_perspectives}
    \end{figure}
    
    Transitioning \acp{CAM} from specialized lookup tables to foundational computing primitives opens the door to high-performance and paradigm-shifting capabilities. To realize this potential, we propose here directions for future work for researchers across many disciplines, summarized in Fig.~\ref{fig:core_future_perspectives}.
    
    \begin{enumerate}
        \item \textbf{Device-Level:} Future \ac{CAM} systems must be co-designed with memory devices that balance density, endurance, retention, analog programmability, and read energy. Unlike conventional \ac{RAM} or \ac{CIM}, many \ac{CAM} workloads can benefit from high-resistance states, since massively parallel search favors low-current readout and reduced static power. Emerging non-volatile devices with high endurance and precise analog programmability therefore offering a promising substrate for energy-efficient associative arrays. To overcome the area penalty of \acp{CAM}, monolithic 3D integration is a key frontier, with comparison logic in the silicon substrate and dense reference weights, defining the stored keys embedded in 3D stacked layers.
        
        \item \textbf{Architectural:} Future \ac{CAM} research should first aim to invent novel circuit and cell designs that maximize sensing gain while minimizing component count, area, and peripheral overhead. Second, large-scale sparse associative search for databases can be explored more deeply, motivated by the overlap between \ac{CAM}-based retrieval, sparse attention, and hierarchical memory access in the brain. Third, future systems should increasingly develop application-specific \ac{CAM} architectures, as prior work shows that tailoring the associative array to the workload can yield substantial gains in area, latency, energy, and reliability, exemplified in this perspective with the \ac{CAM} and crossbar integration. 
        
        \item \textbf{Emerging Application Primitives:} Future research should expand \ac{CAM} deployments beyond static retrieval toward dynamic, probabilistic, and real-time workloads. One promising direction is to use analog associative search as a transition engine for probabilistic or non-deterministic computing, where state transitions are selected not by exact lookup, but by similarity, uncertainty, or learned matching boundaries. This could provide a hardware-native route toward probabilistic CAM-based computing. Another important direction is the use of ultra-fast associative lookup in high-rate scientific workloads, such as real-time particle-track triggering in nuclear physics, where relevant patterns must be identified under strict latency and energy constraints.

        \item \textbf{Machine Learning with \acp{CAM}:} Machine learning with \acp{CAM} remains a niche topic, partly because \ac{AI} researchers rarely view \acp{CAM} as trainable hardware primitives. Future research should investigate efficient surrogate models and online learning methods that make \acp{CAM} easier to integrate into existing machine-learning pipelines. In parallel, in-hardware learning should be explored more deeply, especially forward-only or local learning algorithms that reduce the need for full backpropagation through associative memory arrays.
    
        \item \textbf{Translating Neuroscience into \ac{CAM} Hardware Primitives:} An open and long-term challenge is to translate the brain's ability to retrieve, reorganize, consolidate, and forget information into practical associative hardware. Rather than mimicking biological wetware directly, future \ac{CAM} research should abstract high-level principles such as sparse and hierarchical memory organization, context-dependent recall, surprise-triggered writes, and offline memory consolidation. This points toward dynamic, self-organizing \ac{CAM} arrays that cluster related memories, prune rarely accessed data, store only novel inputs, and replay important patterns during idle phases to migrate information across an associative memory hierarchy.

    \end{enumerate}

    \section*{Acknowledgements}
    We wish to thank Walter Senn for insightful feedback on neuroscientific topics.  We also thank Sebastian Siegel and Maxime Fabre for their feedback and careful reading. This work was supported in part by the Federal Ministry of Research, Technology and Space (BMFTR) through NEUROTEC II, project number 16ME0398K.
    
    \bibliography{literature}

@article{leroux2025analog,
  author       = {Leroux, Nathan and Manea, Paul-Philipp and Sudarshan, Chirag and Finkbeiner, Jan and Siegel, Sebastian and Strachan, John Paul and Neftci, Emre},
  title        = {Analog in-memory computing attention mechanism for fast and energy-efficient large language models},
  journal      = {Nature Computational Science},
  year         = {2025},
  volume       = {5},
  pages        = {544--556},
  doi          = {10.1038/s43588-025-00854-1},
  url          = {https://www.nature.com/articles/s43588-025-00854-1}
}

@inproceedings{10185398,
 author = {Belmonte, A. and Kundu, S. and Subhechha, S. and Chasin, A. and Rassoul, N. and Dekkers, H. and Puliyalil, H. and Seidel, F. and Carolan, P. and Delhougne, R. and Kar, G. S.},
 booktitle = {2023 IEEE Symposium on VLSI Technology and Circuits (VLSI Technology and Circuits)},
 doi = {10.23919/VLSITechnologyandCir57934.2023.10185398},
 pages = {1-2},
 title = {Lowest IOFF \(<\) 3×10-21 A/\textmu m in capacitorless DRAM achieved by Reactive Ion Etch of IGZO-TFT},
 year = {2023}
}

@article{10268258,
 author = {Tang, Wenjun and Liu, Jialong and Sun, Chen and Zheng, Zijie and Liu, Yongpan and Yang, Huazhong and Jiang, Chen and Ni, Kai and Gong, Xiao and Li, Xueqing},
 doi = {10.1109/TCSI.2023.3317170},
 journal = {IEEE Transactions on Circuits and Systems I: Regular Papers},
 number = {12},
 pages = {5166-5179},
 title = {Low-Power and Scalable BEOL-Compatible IGZO TFT eDRAM-Based Charge-Domain Computing},
 volume = {70},
 year = {2023}
}

@inproceedings{9401798,
 author = {Raman, Siddhartha Raman Sundara and Xie, Shanshan and P.Kulkarni, Jaydeep},
 booktitle = {2021 IEEE International Symposium on Circuits and Systems (ISCAS)},
 doi = {10.1109/ISCAS51556.2021.9401798},
 number = {},
 pages = {1-5},
 title = {Compute-in-eDRAM with Backend Integrated Indium Gallium Zinc Oxide Transistors},
 volume = {},
 year = {2021}
}

@article{Goode2020IntegratedIndex,
  author  = {Goode, Travis D. and Tanaka, Kazuhiro Z. and Sahay, Amar and McHugh, Thomas J.},
  title   = {An Integrated Index: Engrams, Place Cells, and Hippocampal Memory},
  journal = {Neuron},
  year    = {2020},
  volume  = {107},
  number  = {5},
  pages   = {805--820},
  doi     = {10.1016/j.neuron.2020.07.011},
  pmid    = {32763146},
  pmcid   = {PMC7486247}
}

@misc{granier2025multiheadselfattentioncorticothalamiccircuits,
      title={Multihead self-attention in cortico-thalamic circuits}, 
      author={Arno Granier and Walter Senn},
      year={2025},
      eprint={2504.06354},
      archivePrefix={arXiv},
      primaryClass={q-bio.NC},
      url={https://arxiv.org/abs/2504.06354}, 
}

@ARTICLE{Kohonen1972,
  author={Kohonen, Teuvo},
  journal={IEEE Transactions on Computers}, 
  title={Correlation Matrix Memories}, 
  year={1972},
  volume={C-21},
  number={4},
  pages={353-359},
  doi={10.1109/TC.1972.5008975}}

@inproceedings{dube2025analog,
 author = {Aradhana Dube and Paul Manea and Paolo Gibertini and Erika Covi and John Paul Strachan},
 booktitle = {Proceedings of the IEEE International Symposium on Circuits and Systems (ISCAS)},
 title = {Analog Softmax with Wide Input Current Range for In-Memory Computing},
 year = {2025}
}

@article{lu2024high,
 author = {Lu, Anni and Lee, Junmo and Kim, Tae-Hyeon and Karim, Muhammed Ahosan Ul and Park, Rebecca Sejung and Simka, Harsono and Yu, Shimeng},
 journal = {Nature Reviews Electrical Engineering},
 number = {1},
 pages = {24--34},
 publisher = {Nature Publishing Group UK London},
 title = {High-speed emerging memories for AI hardware accelerators},
 volume = {1},
 year = {2024}
}

@article{Matsunaga2012A3U,
  title={A 3.14 um2 4T-2MTJ-cell fully parallel TCAM based on nonvolatile logic-in-memory architecture},
  author={Shoun Matsunaga and Sadahiko Miura and Hiroaki Honjo and Keizo Kinoshita and Shoji Ikeda and Tetsuo Endoh and Hideo Ohno and Takahiro Hanyu},
  journal={2012 Symposium on VLSI Circuits (VLSIC)},
  year={2012},
  pages={44-45},
  url={https://api.semanticscholar.org/CorpusID:3449145}
}

@article{Masoodur2t2rtcam,
author = {Masoodur Rahman Khan  and ABM Harun‐ur Rashid },
title = {Memristor‐transistor hybrid ternary content addressable memory using ternary memristive memory cell},
journal = {IET Circuits, Devices \& Systems},
volume = {15},
issue = {7},
pages = {619-629},
year = {2021},
doi = {10.1049/cds2.12057},

URL = {https://digital-library.theiet.org/doi/abs/10.1049/cds2.12057},
eprint = {https://digital-library.theiet.org/doi/pdf/10.1049/cds2.12057}
}

@ARTICLE{8320858,
  author={Mourgias-Alexandris, G. and Vagionas, C. and Tsakyridis, A. and Maniotis, P. and Pleros, N.},
  journal={IEEE Photonics Technology Letters}, 
  title={Optical Content Addressable Memory Matchline for 2-bit Address Look-Up at 10 Gb/s}, 
  year={2018},
  volume={30},
  number={9},
  pages={809-812},
  doi={10.1109/LPT.2018.2817928}}

@INPROCEEDINGS{11041521,
  author={Gao, Guoyun and Wen, Bo and Yang, Ni and Du, Zhiyuan and Jiang, Mingrui and Mao, Ruibin and Cao, Yingnan and Xue, Hongxia and Yip, Pak San and Liu, Qihan and Ki, Dong-Keun and Tang, Jinyao and Chan, Paddy K. L. and Jiang, Hao and Wang, Han and Li, Lain-Jong and Li, Can},
  booktitle={2025 9th IEEE Electron Devices Technology \& Manufacturing Conference (EDTM)}, 
  title={Enabling Highly-Efficient, Low-Latency Analog CAM Operations with Optimized MoS2 Flash Memory Devices}, 
  year={2025},
  volume={},
  number={},
  pages={1-3},
  doi={10.1109/EDTM61175.2025.11041521}}

@ARTICLE{10521670,
  author={Li, Chao and Sun, Chen and Yang, Jianyi and Ni, Kai and Gong, Xiao and Zhuo, Cheng and Yin, Xunzhao},
  journal={IEEE Transactions on Very Large Scale Integration (VLSI) Systems}, 
  title={Multibit Content Addressable Memory Design and Optimization Based on 3-D nand-Compatible IGZO Flash}, 
  year={2024},
  volume={32},
  number={8},
  pages={1380-1388},
  doi={10.1109/TVLSI.2024.3392159}}

@ARTICLE{7894184,
  author={Chang, Meng-Fan and Lin, Chien-Chen and Lee, Albert and Chiang, Yen-Ning and Kuo, Chia-Chen and Yang, Geng-Hau and Tsai, Hsiang-Jen and Chen, Tien-Fu and Sheu, Shyh-Shyuan},
  journal={IEEE Journal of Solid-State Circuits}, 
  title={A 3T1R Nonvolatile TCAM Using MLC ReRAM for Frequent-Off Instant-On Filters in IoT and Big-Data Processing}, 
  year={2017},
  volume={52},
  number={6},
  pages={1664-1679},
  doi={10.1109/JSSC.2017.2681458}}

@ARTICLE{9106766,
  author={Yin, Xunzhao and Li, Chao and Huang, Qingrong and Zhang, Li and Niemier, Michael and Hu, Xiaobo Sharon and Zhuo, Cheng and Ni, Kai},
  journal={IEEE Transactions on Electron Devices}, 
  title={FeCAM: A Universal Compact Digital and Analog Content Addressable Memory Using Ferroelectric}, 
  year={2020},
  volume={67},
  number={7},
  pages={2785-2792},
  doi={10.1109/TED.2020.2994896}}

@Misc{Bazzi2022,
  author     = {Bazzi, Jinane and Sweidan, Jana and Fouda, Mohammed E. and Kanj, Rouwaida and Eltawil, Ahmed M.},
  title      = {Efficient Analog CAM Design},
  year       = {2022},
  comment    = {Their main issue seems to be the gain of the switches, which pull the ML. This will increase the resolutions since the "forbidden Area" will get smaller. However, aren't we losing the differentiable ability through this, since the boundaries get sharper. But I Admit that this is a different thing than the storage capacity (number of possible Intervals).
In the 8T2M design, they replaced the T1 with a PMOS transistor. Isn't that a problem, since PMOS switches usually have poorer electrical properties, and it will most probably have different transients than the other switch T2. At least in my simulations, I could observe this. However, it could also be a simulation issue. That also makes our desired dCAM design problematic, since we want defined and symmetric transients.
Their algorithm to define the resolution is interesting. However, it is just suitable for aCAM and not for dCAM. I have to admit that I did not understand it to 100%, but I will get back to that in case we need something similar.},
  copyright  = {arXiv.org perpetual, non-exclusive license},
  doi        = {10.48550/ARXIV.2203.02500},
  publisher  = {arXiv},
  ranking    = {rank5},
  readstatus = {skimmed},
  url        = {https://arxiv.org/abs/2203.02500},
}

@inproceedings{shi_counteractive_2023,
 author = {Shi, Mingcheng and Su, Yanbo and Tang, Jianshi and Li, Yijun and Du, Yiwei and An, Ran and Li, Jiaming and Li, Yuankun and Yao, Jian and Hu, Ruofei and He, Yuan and Xi, Yue and Li, Qingwen and Qiu, Song and Zhang, Qingtian and Pan, Liyang and Gao, Bin and Qian, He and Wu, Huaqiang},
 booktitle = {2023 {International} {Electron} {Devices} {Meeting} ({IEDM})},
 doi = {10.1109/IEDM45741.2023.10413876},
 month = {December},
 pages = {1--4},
 title = {Counteractive {Coupling} {IGZO}/{CNT} {Hybrid} {2T0C} {DRAM} {Accelerating} {RRAM}-based {Computing}-{In}-{Memory} via {Monolithic} {3D} {Integration} for {Edge} {AI}},
 url = {https://ieeexplore.ieee.org/document/10413876},
 year = {2023}
}

@inproceedings{vaswani2023attentionneed,
	address = {Red Hook, NY, USA},
	series = {{NIPS}'17},
	title = {Attention is all you need},
	isbn = {978-1-5108-6096-4},
	booktitle = {Proceedings of the 31st {International} {Conference} on {Neural} {Information} {Processing} {Systems}},
	publisher = {Curran Associates Inc.},
	author = {Vaswani, Ashish and Shazeer, Noam and Parmar, Niki and Uszkoreit, Jakob and Jones, Llion and Gomez, Aidan N. and Kaiser, {\L{}}ukasz and Polosukhin, Illia},
	month = dec,
	year = {2017},
	pages = {6000--6010},
}

@misc{vogginger2024neuromorphichardwaresustainableai,
 archiveprefix = {arXiv},
 author = {Bernhard Vogginger and Amirhossein Rostami and Vaibhav Jain and Sirine Arfa and Andreas Hantsch and David Kappel and Michael Schäfer and Ulrike Faltings and Hector A. Gonzalez and Chen Liu and Christian Mayr and Wolfgang Maaß},
 eprint = {2402.02521},
 primaryclass = {cs.ET},
 title = {Neuromorphic hardware for sustainable AI data centers},
 url = {https://arxiv.org/abs/2402.02521},
 year = {2024}
}

@inproceedings{ye_double-gate_2020,
 author = {Ye, H. and Gomez, J. and Chakraborty, W. and Spetalnick, S. and Dutta, S. and Ni, K. and Raychowdhury, A. and Datta, S.},
 booktitle = {2020 {IEEE} {International} {Electron} {Devices} {Meeting} ({IEDM})},
 doi = {10.1109/IEDM13553.2020.9371981},
 month = {December},
 pages = {28.3.1--28.3.4},
 title = {Double-{Gate} {W}-{Doped} {Amorphous} {Indium} {Oxide} {Transistors} for {Monolithic} {3D} {Capacitorless} {Gain} {Cell} {eDRAM}},
 url = {https://ieeexplore.ieee.org/document/9371981/?arnumber=9371981},
 year = {2020}
}

@inproceedings{manea2023non,
  title={Non-idealities and design solutions for analog memristor-based content-addressable memories},
  author={Manea, Paul-Philipp and Sudarshan, Chirag and C{\"u}ppers, Felix and Strachan, John Paul},
  booktitle={Proceedings of the 18th ACM International Symposium on Nanoscale Architectures},
  pages={1--6},
  year={2023}
}

@INPROCEEDINGS{maena_gain_cell_aCAM,
  author={Manea, Paul-Philipp and Leroux, Nathan and Neftci, Emre and Strachan, John Paul},
  booktitle={2025 IEEE International Symposium on Circuits and Systems (ISCAS)}, 
  title={Gain Cell-based Analog Content Addressable Memory for Dynamic Associative Tasks in AI}, 
  year={2025},
  volume={},
  number={},
  pages={1-5},
  doi={10.1109/ISCAS56072.2025.11044190}}

@misc{zhao2025nldpeanaloginmemorynonlinear,
      title={NL-DPE: An Analog In-memory Non-Linear Dot Product Engine for Efficient CNN and LLM Inference}, 
      author={Lei Zhao and Luca Buonanno and Archit Gajjar and John Moon and Aishwarya Natarajan and Sergey Serebryakov and Ron M. Roth and Xia Sheng and Youtao Zhang and Paolo Faraboschi and Jim Ignowski and Giacomo Pedretti},
      year={2025},
      eprint={2511.13950},
      archivePrefix={arXiv},
      primaryClass={cs.AR},
      url={https://arxiv.org/abs/2511.13950}, 
}

@article{shafiee2016isaac,
  title={ISAAC: A convolutional neural network accelerator with in-situ analog arithmetic in crossbars},
  author={Shafiee, Ali and Nag, Anirban and Muralimanohar, Naveen and Balasubramonian, Rajeev and Strachan, John Paul and Hu, Miao and Williams, R Stanley and Srikumar, Vivek},
  journal={ACM SIGARCH Computer Architecture News},
  volume={44},
  number={3},
  pages={14--26},
  year={2016},
  publisher={ACM New York, NY, USA}
}

@article{graves2016differentiable,
  title={Hybrid computing using a neural network with dynamic external memory},
  author={Graves, Alex and Wayne, Greg and Reynolds, Malcolm and Harley, Tim and Danihelka, Ivo and Grabska-Barwi{\'n}ska, Agnieszka and Gomez Colmenarejo, Sergio and Grefenstette, Edward and Ramalho, Tiago and Agapiou, John and Badia, Adri{\`a} Puigdom{\`e}nech and Hermann, Karl Moritz and Zwols, Yori and Ostrovski, Georg and Cain, Adam and King, Helen and Summerfield, Christopher and Blunsom, Phil and Kavukcuoglu, Koray and Hassabis, Demis},
  journal={Nature},
  volume={538},
  number={7626},
  pages={471--476},
  year={2016},
  publisher={Nature Publishing Group},
  doi={10.1038/nature20101},
  url={https://www.nature.com/articles/nature20101}
}

@Article{1599540,
  author   = {Pagiamtzis, K. and Sheikholeslami, A.},
  journal  = {IEEE Journal of Solid-State Circuits},
  title    = {Content-addressable memory (CAM) circuits and architectures: a tutorial and survey},
  year     = {2006},
  number   = {3},
  pages    = {712-727},
  volume   = {41},
  doi      = {10.1109/JSSC.2005.864128},
}

@misc{zhong2025understandingtransformerperspectiveassociative,
      title={Understanding Transformer from the Perspective of Associative Memory}, 
      author={Shu Zhong and Mingyu Xu and Tenglong Ao and Guang Shi},
      year={2025},
      eprint={2505.19488},
      archivePrefix={arXiv},
      primaryClass={cs.LG},
      url={https://arxiv.org/abs/2505.19488}, 
}

@article{bietti2023birth,
  title={Birth of a transformer: A memory viewpoint},
  author={Bietti, Alberto and Cabannes, Vivien and Bouchacourt, Diane and Jegou, Herve and Bottou, Leon},
  journal={Advances in Neural Information Processing Systems},
  volume={36},
  pages={1560--1588},
  year={2023}
}

@misc{chalkidis2022explorationhierarchicalattentiontransformers,
      title={An Exploration of Hierarchical Attention Transformers for Efficient Long Document Classification}, 
      author={Ilias Chalkidis and Xiang Dai and Manos Fergadiotis and Prodromos Malakasiotis and Desmond Elliott},
      year={2022},
      eprint={2210.05529},
      archivePrefix={arXiv},
      primaryClass={cs.CL},
      url={https://arxiv.org/abs/2210.05529}, 
}

@article{roy2021efficient,
  title={Efficient content-based sparse attention with routing transformers},
  author={Roy, Aurko and Saffar, Mohammad and Vaswani, Ashish and Grangier, David},
  journal={Transactions of the Association for Computational Linguistics},
  volume={9},
  pages={53--68},
  year={2021}
}

@misc{child2019generatinglongsequencessparse,
      title={Generating Long Sequences with Sparse Transformers}, 
      author={Rewon Child and Scott Gray and Alec Radford and Ilya Sutskever},
      year={2019},
      eprint={1904.10509},
      archivePrefix={arXiv},
      primaryClass={cs.LG},
      url={https://arxiv.org/abs/1904.10509}, 
}

@misc{molomochir2025camformerassociativememoryneed,
  title         = {CAMformer: Associative Memory is All You Need},
  author        = {Tergel Molom-Ochir and Benjamin F. Morris and Mark Horton and Chiyue Wei and Cong Guo and Brady Taylor and Peter Liu and Shan X. Wang and Deliang Fan and Hai Helen Li and Yiran Chen},
  year          = {2025},
  eprint        = {2511.19740},
  archivePrefix = {arXiv},
  primaryClass  = {cs.AR},
  url           = {https://arxiv.org/abs/2511.19740}
}

@article{Hong2025AdaptiveADC,
  author    = {Hong, Haiqiao and Du, Zhiyuan and Jiang, Mingrui and Mao, Ruibin and Ren, Yuan and Li, Fuyi and Mao, Wei and Peng, Muyuan and Zhang, Wei and Liu, Zhengwu and Li, Can and Wong, Ngai},
  title     = {Memristor-based Adaptive Analog-to-Digital Conversion for Efficient and Accurate Compute-in-Memory},
  journal   = {Nature Communications},
  volume    = {16},
  number    = {1},
  pages     = {9749},
  year      = {2025},
  month     = nov,
  publisher = {Nature Publishing Group UK},
  doi       = {10.1038/s41467-025-49749-7}
}

@INPROCEEDINGS{9372119,
  author={Li, Chao and Müller, Franz and Ali, Tarek and Olivo, Ricardo and Imani, Mohsen and Deng, Shan and Zhuo, Cheng and Kämpfe, Thomas and Yin, Xunzhao and Ni, Kai},
  booktitle={2020 IEEE International Electron Devices Meeting (IEDM)}, 
  title={A Scalable Design of Multi-Bit Ferroelectric Content Addressable Memory for Data-Centric Computing}, 
  year={2020},
  volume={},
  number={},
  pages={29.3.1-29.3.4},
  doi={10.1109/IEDM13553.2020.9372119}}

@ARTICLE{11150515,
  author={Liu, Liu and Mehdi Sharifi, Mohammad and Wang, Kunshi and Mao, Ruibin and Ni, Kai and Li, Can and Yin, Xunzhao and Niemier, Michael and Sharon Hu, Xiaobo},
  journal={IEEE Transactions on Computer-Aided Design of Integrated Circuits and Systems}, 
  title={EvaCAM: A Circuit-Level Evaluation Tool for General Content Addressable Memories}, 
  year={2026},
  volume={45},
  number={4},
  pages={1663-1676},
  doi={10.1109/TCAD.2025.3605857}}

@ARTICLE{10753423,
  author={Pedretti, Giacomo and Moon, John and Bruel, Pedro and Serebryakov, Sergey and Roth, Ron M. and Buonanno, Luca and Gajjar, Archit and Zhao, Lei and Ziegler, Tobias and Xu, Cong and Foltin, Martin and Faraboschi, Paolo and Ignowski, Jim and Graves, Catherine E.},
  journal={IEEE Journal on Exploratory Solid-State Computational Devices and Circuits}, 
  title={X-TIME: Accelerating Large Tree Ensembles Inference for Tabular Data With Analog CAMs}, 
  year={2024},
  volume={10},
  number={},
  pages={116-124},
  doi={10.1109/JXCDC.2024.3495634}}

@article{Zhao2023RACEITAR,
  title={RACE-IT: A Reconfigurable Analog CAM-Crossbar Engine for In-Memory Transformer Acceleration},
  author={Lei Zhao and Luca Buonanno and Ron M. Roth and Sergey Serebryakov and Archit Gajjar and John Moon and Jim Ignowski and Giacomo Pedretti},
  journal={ArXiv},
  year={2023},
  volume={abs/2312.06532},
  url={https://api.semanticscholar.org/CorpusID:266162489}
}

@ARTICLE{5219279,
  author={Frei, E. H. and Goldberg, J.},
  journal={IRE Transactions on Electronic Computers}, 
  title={A Method for Resolving Multiple Responses in a Parallel Search File}, 
  year={1961},
  volume={EC-10},
  number={4},
  pages={718-722},
  doi={10.1109/TEC.1961.5219279}}

@inproceedings{Kilburn1961AtlasSupervisor,
  author    = {Tom Kilburn and R. B. Payne and D. J. Howarth},
  title     = {The Atlas Supervisor},
  booktitle = {Proceedings of the December 12--14, 1961 Eastern Joint Computer Conference: Computers --- Key to Total Systems Control},
  series    = {AFIPS '61 (Eastern)},
  pages     = {279--294},
  year      = {1961},
  publisher = {ACM},
  doi       = {10.1145/1460764.1460786},
}

@ARTICLE{34085,
  author={Wade, J.P. and Sodini, C.G.},
  journal={IEEE Journal of Solid-State Circuits}, 
  title={A ternary content addressable search engine}, 
  year={1989},
  volume={24},
  number={4},
  pages={1003-1013},
  doi={10.1109/4.34085}}

@manual{Intel1973C3104CAM,
  author       = {{Intel Corporation}},
  title        = {3104 High-Speed 16-Bit Content Addressable Memory},
  year         = {1973},
  url          = {https://deramp.com/downloads/mfe_archive/050-Component%20Specifications/Intel/Memory%20Components/Intel%20C3104%20CAM.pdf},
  note         = {Intel C3104 CAM datasheet. Accessed: 2026-03-12}
}

@inproceedings{Yamagata1991StackedCapacitorCAM,
  author    = {Tadato Yamagata and Masaaki Mihara and Takeshi Hamamoto and Toshifumi Kobayashi and Michihiro Yamada},
  title     = {A 288-kbit Fully Parallel Content Addressable Memory Using Stacked Capacitor Cell Structure},
  booktitle = {Proceedings of the IEEE 1991 Custom Integrated Circuits Conference (CICC)},
  year      = {1991},
  pages     = {138--141},
  address   = {San Diego, CA, USA},
  publisher = {IEEE},
  doi       = {10.1109/CICC.1991.164123}
}

@mastersthesis{Lee1993InexactCAM,
  author  = {Jack Lee},
  title   = {Smart Memory: An Inexact Content-Addressable Memory},
  school  = {Portland State University},
  year    = {1993},
  type    = {Master's Thesis},
  doi     = {10.15760/etd.6489},
  url     = {https://pdxscholar.library.pdx.edu/cgi/viewcontent.cgi?article=5676&context=open_access_etds}
}

@inproceedings{zane2003coolcams,
  title={CoolCAMs: Power-efficient TCAMs for forwarding engines},
  author={Zane, Francis and Narlikar, Girija and Basu, Anindya},
  booktitle={IEEE INFOCOM 2003. Twenty-second Annual Joint Conference of the IEEE Computer and Communications Societies (IEEE Cat. No. 03CH37428)},
  volume={1},
  pages={42--52},
  year={2003},
  organization={IEEE}
}

@inproceedings{mcauley1993fast,
  title={Fast routing table lookup using CAMs},
  author={McAuley, Anthony J and Francis, Paul},
  booktitle={IEEE INFOCOM'93 The Conference on Computer Communications, Proceedings},
  pages={1382--1391},
  year={1993},
  organization={IEEE}
}

@ARTICLE{5471063,
  author={Eshraghian, Kamran and Cho, Kyoung-Rok and Kavehei, Omid and Kang, Soon-Ku and Abbott, Derek and Kang, Sung-Mo Steve},
  journal={IEEE Transactions on Very Large Scale Integration (VLSI) Systems}, 
  title={Memristor MOS Content Addressable Memory (MCAM): Hybrid Architecture for Future High Performance Search Engines}, 
  year={2011},
  volume={19},
  number={8},
  pages={1407-1417},
  doi={10.1109/TVLSI.2010.2049867}}

@INPROCEEDINGS{130700,
  author={Hanyu, T. and Higuchi, T.},
  booktitle={[1991] Proceedings of the Twenty-First International Symposium on Multiple-Valued Logic}, 
  title={A floating-gate-MOS-based multiple-valued associative memory}, 
  year={1991},
  volume={},
  number={},
  pages={24-31},
  doi={10.1109/ISMVL.1991.130700}}

@article{mathan2016data,
  title={Data intelligent low power high performance TCAM for IP-address lookup table},
  author={Mathan, K and Ravichandran, T},
  journal={Circuits Syst},
  volume={7},
  number={11},
  pages={3734},
  year={2016}
}

@ARTICLE{Hybrid_CAM_SRAM_sparse_attention,
  author={Huang, Haiqiu and Wang, Mingyu and Li, Xiaojie and Zhong, Baiqing and Yang, Zeqi and Lu, Tao and Zhang, Yicong and Yu, Zhiyi},
  journal={IEEE Transactions on Circuits and Systems II: Express Briefs}, 
  title={A Hybrid CAM-SRAM Processing-in-Memory Architecture With Feature Level Sparsity for Attention Mechanisms}, 
  year={2025},
  volume={72},
  number={9},
  pages={1283-1287},
  doi={10.1109/TCSII.2025.3590432}}

@INPROCEEDINGS{6858404,
  author={Li-Yue Huang and Meng-Fan Chang and Ching-Hao Chuang and Chia-Chen Kuo and Chien-Fu Chen and Geng-Hau Yang and Hsiang-Jen Tsai and Tien-Fu Chen and Shyh-Shyuan Sheu and Keng-Li Su and Chen, Frederick T. and Tzu-Kun Ku and Ming-Jinn Tsai and Ming-Jer Kao},
  booktitle={2014 Symposium on VLSI Circuits Digest of Technical Papers}, 
  title={ReRAM-based 4T2R nonvolatile TCAM with 7x NVM-stress reduction, and 4x improvement in speed-wordlength-capacity for normally-off instant-on filter-based search engines used in big-data processing}, 
  year={2014},
  volume={},
  number={},
  pages={1-2},
  doi={10.1109/VLSIC.2014.6858404}}

@INPROCEEDINGS{11040774,
  author={Manea, Paul-Philipp and Leroux, Nathan and Strachan, John Paul},
  booktitle={2025 9th IEEE Electron Devices Technology \& Manufacturing Conference (EDTM)}, 
  title={Content Addressable Memory Hierarchies for Computing in Memory}, 
  year={2025},
  volume={},
  number={},
  pages={1-3},
  doi={10.1109/EDTM61175.2025.11040774}}

@ARTICLE{7563786,
  author={Chang, Meng-Fan and Huang, Lie-Yue and Lin, Wen-Zhang and Chiang, Yen-Ning and Kuo, Chia-Chen and Chuang, Ching-Hao and Yang, Keng-Hao and Tsai, Hsiang-Jen and Chen, Tien-Fu and Sheu, Shyh-Shyuan},
  journal={IEEE Journal of Solid-State Circuits}, 
  title={A ReRAM-Based 4T2R Nonvolatile TCAM Using RC-Filtered Stress-Decoupled Scheme for Frequent-OFF Instant-ON Search Engines Used in IoT and Big-Data Processing}, 
  year={2016},
  volume={51},
  number={11},
  pages={2786-2798},
  doi={10.1109/JSSC.2016.2602218}}

@INPROCEEDINGS{11563370,
  author={Liu, Liu and Pereira, Tomas Sousa and Sharifi, Mohammad Mehdi and Ren, Pengyu and Wen, Bo and Li, Can and Ni, Kai and Niemier, Michael and Hu, X. Sharon},
  booktitle={2026 IEEE 44th VLSI Test Symposium (VTS)}, 
  title={Enhancing Robustness of Content-Addressable Memories for In-Memory Search}, 
  year={2026},
  volume={},
  number={},
  pages={1-5},
  doi={10.1109/VTS69484.2026.11563370}}

@INPROCEEDINGS{10848806,
  author={Yu, Jiaao and Manea, Paul-Philipp and Hizzani, Mohammad and Ameli, Sara and Strachan, John Paul},
  booktitle={2024 31st IEEE International Conference on Electronics, Circuits and Systems (ICECS)}, 
  title={A Memristor Variation-Aware Analog Memristor Programming Circuit for Associative Memories}, 
  year={2024},
  volume={},
  number={},
  pages={1-4},
  doi={10.1109/ICECS61496.2024.10848806}}

@INPROCEEDINGS{7063054,
  author={Chang, Meng-Fan and Lin, Chien-Chen and Lee, Albert and Kuo, Chia-Chen and Yang, Geng-Hau and Tsai, Hsiang-Jen and Chen, Tien-Fu and Sheu, Shyh-Shyuan and Tseng, Pei-Ling and Lee, Heng-Yuan and Ku, Tzu-Kun},
  booktitle={2015 IEEE International Solid-State Circuits Conference - (ISSCC) Digest of Technical Papers}, 
  title={17.5 A 3T1R nonvolatile TCAM using MLC ReRAM with Sub-1ns search time}, 
  year={2015},
  volume={},
  number={},
  pages={1-3},
  doi={10.1109/ISSCC.2015.7063054}}

@misc{Manea2026SALM,
      title={A Fast and Energy-Efficient Latch-Based Memristive Analog Content-Addressable Memory}, 
      author={Paul-Philipp Manea and Aishwarya Natarajan and Jim Ignowski and John Paul Strachan and Luca Buonanno},
      year={2026},
      eprint={2605.11847},
      archivePrefix={arXiv},
      primaryClass={cs.ET},
      url={https://arxiv.org/abs/2605.11847}, 
}

@INPROCEEDINGS{Yu_CAM_LUT_KA,
  author={Yu, Xuliang and Qian, Yu and Yin, Xunzhao and Zhuo, Cheng and Zhao, Liang},
  booktitle={2025 62nd ACM/IEEE Design Automation Conference (DAC)}, 
  title={FeKAN: Efficient Kolmogorov-Arnold Networks Accelerator Using FeFET-based CAM and LUT}, 
  year={2025},
  volume={},
  number={},
  pages={1-7},
  doi={10.1109/DAC63849.2025.11132687}}

@inproceedings{Ankit2019PUMA,
  author    = {Aayush Ankit and Izzat El Hajj and Sai Rahul Chalamalasetti and Geoffrey Ndu and Martin Foltin and R. Stanley Williams and Paolo Faraboschi and Wen-mei Hwu and John Paul Strachan and Kaushik Roy and Dejan S. Milojicic},
  title     = {PUMA: A Programmable Ultra-Efficient Memristor-Based Accelerator for Machine Learning Inference},
  booktitle = {Proceedings of the Twenty-Fourth International Conference on Architectural Support for Programming Languages and Operating Systems (ASPLOS)},
  pages     = {715--731},
  year      = {2019},
  doi       = {10.1145/3297858.3304049}
}

@article{yao2020fully,
  title={Fully hardware-implemented memristor convolutional neural network},
  author={Yao, Peng and Wu, Huaqiang and Gao, Bin and Tang, Jianshi and Zhang, Qingtian and Zhang, Wenqiang and Yang, J Joshua and Qian, He},
  journal={Nature},
  volume={577},
  number={7792},
  pages={641--646},
  year={2020},
  publisher={Nature Publishing Group UK London}
}

@article{Kuon_FPGA_2008,
author = {Kuon, Ian and Tessier, Russell and Rose, Jonathan},
title = {FPGA Architecture: Survey and Challenges},
year = {2008},
issue_date = {Apr 2008},
publisher = {Now Publishers Inc.},
address = {Hanover, MA, USA},
volume = {2},
number = {2},
issn = {1551-3939},
url = {https://doi.org/10.1561/1000000005},
doi = {10.1561/1000000005},
journal = {Found. Trends Electron. Des. Autom.},
month = apr,
pages = {135–253},
numpages = {119}
}

@article{Graves2020FSM,
author = {Graves, Catherine E. and Li, Can and Sheng, Xia and Miller, Darrin and Ignowski, Jim and Kiyama, Lennie and Strachan, John Paul},
title = {In-Memory Computing with Memristor Content Addressable Memories for Pattern Matching},
journal = {Advanced Materials},
volume = {32},
number = {37},
pages = {2003437},
doi = {https://doi.org/10.1002/adma.202003437},
url = {https://advanced.onlinelibrary.wiley.com/doi/abs/10.1002/adma.202003437},
eprint = {https://advanced.onlinelibrary.wiley.com/doi/pdf/10.1002/adma.202003437},
year = {2020}
}

@misc{rezazadeh2025isolatedconversationshierarchicalschemas,
      title={From Isolated Conversations to Hierarchical Schemas: Dynamic Tree Memory Representation for LLMs}, 
      author={Alireza Rezazadeh and Zichao Li and Wei Wei and Yujia Bao},
      year={2025},
      eprint={2410.14052},
      archivePrefix={arXiv},
      primaryClass={cs.CL},
      url={https://arxiv.org/abs/2410.14052}, 
}

@ARTICLE{1327749,
  author={Pagiamtzis, K. and Sheikholeslami, A.},
  journal={IEEE Journal of Solid-State Circuits}, 
  title={A low-power content-addressable memory (CAM) using pipelined hierarchical search scheme}, 
  year={2004},
  volume={39},
  number={9},
  pages={1512-1519},
  doi={10.1109/JSSC.2004.831433},
  ISSN={1558-173X},
  month={Sep.},}

@misc{ramsauer2021hopfieldnetworksneed,
      title={Hopfield Networks is All You Need}, 
      author={Hubert Ramsauer and Bernhard Schäfl and Johannes Lehner and Philipp Seidl and Michael Widrich and Thomas Adler and Lukas Gruber and Markus Holzleitner and Milena Pavlović and Geir Kjetil Sandve and Victor Greiff and David Kreil and Michael Kopp and Günter Klambauer and Johannes Brandstetter and Sepp Hochreiter},
      year={2021},
      eprint={2008.02217},
      archivePrefix={arXiv},
      primaryClass={cs.NE},
      url={https://arxiv.org/abs/2008.02217}, 
}

@Article{Li2020,
  author    = {Can Li and Catherine E. Graves and Xia Sheng and Darrin Miller and Martin Foltin and Giacomo Pedretti and John Paul Strachan},
  journal   = {Nature Communications},
  title     = {Analog content-addressable memories with memristors},
  year      = {2020},
  month     = {apr},
  number    = {1},
  volume    = {11},
  comment   = {100ps pulses
0.52fJ/cell/search},
  doi       = {10.1038/s41467-020-15254-4},
  publisher = {Springer Science and Business Media {LLC}},
  ranking   = {rank5},
}

@article{alkabani2020oe,
  title={OE-CAM: a hybrid opto-electronic content addressable memory},
  author={Alkabani, Yousra and Miscuglio, Mario and Sorger, Volker J and El-Ghazawi, Tarek},
  journal={IEEE Photonics Journal},
  volume={12},
  number={2},
  pages={1--14},
  year={2020},
  publisher={IEEE}
}

@article{ni2019ferroelectric,
  title={Ferroelectric ternary content-addressable memory for one-shot learning},
  author={Ni, Kai and Yin, Xunzhao and Laguna, Ann Franchesca and Joshi, Siddharth and D{\"u}nkel, Stefan and Trentzsch, Martin and M{\"u}ller, Johannes and Beyer, Sven and Niemier, Michael and Hu, Xiaobo Sharon and others},
  journal={Nature Electronics},
  volume={2},
  number={11},
  pages={521--529},
  year={2019},
  publisher={Nature Publishing Group UK London}
}

@Article{Pedretti2021Tree,
  author     = {Giacomo Pedretti and Catherine E. Graves and Sergey Serebryakov and Ruibin Mao and Xia Sheng and Martin Foltin and Can Li and John Paul Strachan},
  journal    = {Nature Communications},
  title      = {Tree-based machine learning performed in-memory with memristive analog {CAM}},
  year       = {2021},
  month      = {oct},
  number     = {1},
  volume     = {12},
  doi        = {10.1038/s41467-021-25873-0},
  publisher  = {Springer Science and Business Media {LLC}},
  ranking    = {rank5},
  readstatus = {read},
}

@Article{Pedretti2022differentiableCAM,
  author    = {Giacomo Pedretti and Catherine E. Graves and Thomas Van Vaerenbergh and Sergey Serebryakov and Martin Foltin and Xia Sheng and Ruibin Mao and Can Li and John Paul Strachan},
  journal   = {Advanced Electronic Materials},
  title     = {Differentiable Content Addressable Memory with Memristors},
  year      = {2022},
  month     = {mar},
  number    = {8},
  pages     = {2101198},
  volume    = {8},
  doi       = {10.1002/aelm.202101198},
  publisher = {Wiley},
}

@ARTICLE{9656637,
  author={Kazemi, Arman and Sharifi, Mohammad Mehdi and Laguna, Ann Franchesca and Müller, Franz and Yin, Xunzhao and Kämpfe, Thomas and Niemier, Michael and Hu, X. Sharon},
  journal={IEEE Transactions on Computers}, 
  title={FeFET Multi-Bit Content-Addressable Memories for In-Memory Nearest Neighbor Search}, 
  year={2022},
  volume={71},
  number={10},
  pages={2565-2576},
  doi={10.1109/TC.2021.3136576}}

@article{gao2025sb,
  title={Sb-contacted MoS2 flash memory for analogue in-memory searches},
  author={Gao, Guoyun and Wen, Bo and Yang, Ni and Du, Zhiyuan and Jiang, Mingrui and Mao, Ruibin and Qiu, Rui and Cao, Yingnan and Xue, Hongxia and Zou, Deng and others},
  journal={Nature Nanotechnology},
  pages={1--9},
  year={2025},
  publisher={Nature Publishing Group UK London}
}

@misc{Manea2022TorchCAM,
  author       = {Paul Manea},
  title        = {TorchCAM},
  year         = {2022},
  month        = may,
  publisher    = {GitLab},
  url          = {https://iffgit.fz-juelich.de/manea/torchcam},
  note         = {GitLab repository, created May 4, 2022. Accessed March 14, 2026}
}

@INPROCEEDINGS{manea2025gaincellcam,
  author={Manea, Paul-Philipp and Leroux, Nathan and Neftci, Emre and Strachan, John Paul},
  booktitle={2025 IEEE International Symposium on Circuits and Systems (ISCAS)}, 
  title={Gain Cell-based Analog Content Addressable Memory for Dynamic Associative Tasks in AI}, 
  year={2025},
  volume={},
  number={},
  pages={1-5},
  doi={10.1109/ISCAS56072.2025.11044190}}

@article{gershman2025key,
  title={Key-value memory in the brain},
  author={Gershman, Samuel J and Fiete, Ila and Irie, Kazuki},
  journal={Neuron},
  volume={113},
  number={11},
  pages={1694--1707},
  year={2025},
  publisher={Elsevier}
}

@misc{cornell2025hierarchical,
  title={Hierarchical Context Guides Human Memory Search},
  author={Cornell, Charlotte and Zhang, Qiong},
  year={2025},
  month=oct,
  howpublished={PsyArXiv preprint},
  publisher={OSF},
  url={https://doi.org/10.31234/osf.io/mbwe9_v1}
}

@article {Peer2025CognitiveMaps,
	author = {Peer, Michael and Epstein, Russell A.},
	title = {Cognitive maps for hierarchical spaces in the human brain},
	elocation-id = {2025.02.05.636580},
	year = {2025},
	doi = {10.1101/2025.02.05.636580},
	publisher = {Cold Spring Harbor Laboratory},
	URL = {https://www.biorxiv.org/content/early/2025/02/05/2025.02.05.636580},
	eprint = {https://www.biorxiv.org/content/early/2025/02/05/2025.02.05.636580.full.pdf},
	journal = {bioRxiv}
}

@article{teyler1986hippocampal,
  title={The hippocampal memory indexing theory},
  author={Teyler, Timothy J. and DiScenna, Pascal},
  journal={Behavioral Neuroscience},
  volume={100},
  number={2},
  pages={147--154},
  year={1986},
  month={apr},
  publisher={American Psychological Association},
  doi={10.1037/0735-7044.100.2.147},
  pmid={3008780}
}

@article{lin2023recent,
  title={Recent advances in 2D material theory, synthesis, properties, and applications},
  author={Lin, Yu-Chuan and Torsi, Riccardo and Younas, Rehan and Hinkle, Christopher L and Rigosi, Albert F and Hill, Heather M and Zhang, Kunyan and Huang, Shengxi and Shuck, Christopher E and Chen, Chen and others},
  journal={ACS nano},
  volume={17},
  number={11},
  pages={9694--9747},
  year={2023},
  publisher={ACS Publications}
}

@article{liu2021promises,
  title={Promises and prospects of two-dimensional transistors},
  author={Liu, Yuan and Duan, Xidong and Shin, Hyeon-Jin and Park, Seongjun and Huang, Yu and Duan, Xiangfeng},
  journal={Nature},
  volume={591},
  number={7848},
  pages={43--53},
  year={2021},
  publisher={Nature Publishing Group UK London}
}

@article{lu2024projected,
  title={Projected performance of Si-and 2D-material-based SRAM circuits ranging from 16 nm to 1 nm technology nodes},
  author={Lu, Yu-Cheng and Huang, Jing-Kai and Chao, Kai-Yuan and Li, Lain-Jong and Hu, Vita Pi-Ho},
  journal={Nature Nanotechnology},
  volume={19},
  number={7},
  pages={1066--1072},
  year={2024},
  publisher={Nature Publishing Group UK London}
}

@article{liu2021ultrafast,
  title={Ultrafast non-volatile flash memory based on van der Waals heterostructures},
  author={Liu, Lan and Liu, Chunsen and Jiang, Lilai and Li, Jiayi and Ding, Yi and Wang, Shuiyuan and Jiang, Yu-Gang and Sun, Ya-Bin and Wang, Jianlu and Chen, Shiyou and others},
  journal={Nature nanotechnology},
  volume={16},
  number={8},
  pages={874--881},
  year={2021},
  publisher={Nature Publishing Group UK London}
}

@article{yu2023simultaneously,
  title={Simultaneously ultrafast and robust two-dimensional flash memory devices based on phase-engineered edge contacts},
  author={Yu, Jun and Wang, Han and Zhuge, Fuwei and Chen, Zirui and Hu, Man and Xu, Xiang and He, Yuhui and Ma, Ying and Miao, Xiangshui and Zhai, Tianyou},
  journal={Nature Communications},
  volume={14},
  number={1},
  pages={5662},
  year={2023},
  publisher={Nature Publishing Group UK London}
}

@article{wu2021atomically,
  title={Atomically sharp interface enabled ultrahigh-speed non-volatile memory devices},
  author={Wu, Liangmei and Wang, Aiwei and Shi, Jinan and Yan, Jiahao and Zhou, Zhang and Bian, Ce and Ma, Jiajun and Ma, Ruisong and Liu, Hongtao and Chen, Jiancui and others},
  journal={Nature nanotechnology},
  volume={16},
  number={8},
  pages={882--887},
  year={2021},
  publisher={Nature Publishing Group UK London}
}

@article{jiang2024scalable,
  title={A scalable integration process for ultrafast two-dimensional flash memory},
  author={Jiang, Yongbo and Liu, Chunsen and Cao, Zhenyuan and Li, Chuhang and Liu, Zizheng and Wang, Chong and Xiang, Yutong and Zhou, Peng},
  journal={Nature Electronics},
  volume={7},
  number={10},
  pages={868--875},
  year={2024},
  publisher={Nature Publishing Group UK London}
}

@article{das2013high,
  title={High performance multilayer MoS2 transistors with scandium contacts},
  author={Das, Saptarshi and Chen, Hong-Yan and Penumatcha, Ashish Verma and Appenzeller, Joerg},
  journal={Nano letters},
  volume={13},
  number={1},
  pages={100--105},
  year={2013},
  publisher={ACS Publications}
}

@article{liu2022fermi,
  title={Fermi level pinning dependent 2D semiconductor devices: challenges and prospects},
  author={Liu, Xiaochi and Choi, Min Sup and Hwang, Euyheon and Yoo, Won Jong and Sun, Jian},
  journal={Advanced Materials},
  volume={34},
  number={15},
  pages={2108425},
  year={2022},
  publisher={Wiley Online Library}
}

@article{li2023approaching,
  title={Approaching the quantum limit in two-dimensional semiconductor contacts},
  author={Li, Weisheng and Gong, Xiaoshu and Yu, Zhihao and Ma, Liang and Sun, Wenjie and Gao, Si and K{\"o}ro{\u{g}}lu, {\c{C}}a{\u{g}}{\i}l and Wang, Wenfeng and Liu, Lei and Li, Taotao and others},
  journal={Nature},
  volume={613},
  number={7943},
  pages={274--279},
  year={2023},
  publisher={Nature Publishing Group UK London}
}

@article{shen2021ultralow,
  title={Ultralow contact resistance between semimetal and monolayer semiconductors},
  author={Shen, Pin-Chun and Su, Cong and Lin, Yuxuan and Chou, Ang-Sheng and Cheng, Chao-Ching and Park, Ji-Hoon and Chiu, Ming-Hui and Lu, Ang-Yu and Tang, Hao-Ling and Tavakoli, Mohammad Mahdi and others},
  journal={Nature},
  volume={593},
  number={7858},
  pages={211--217},
  year={2021},
  publisher={Nature Publishing Group UK London}
}

@article{li2021epitaxial,
  title={Epitaxial growth of wafer-scale molybdenum disulfide semiconductor single crystals on sapphire},
  author={Li, Taotao and Guo, Wei and Ma, Liang and Li, Weisheng and Yu, Zhihao and Han, Zhen and Gao, Si and Liu, Lei and Fan, Dongxu and Wang, Zixuan and others},
  journal={Nature Nanotechnology},
  volume={16},
  number={11},
  pages={1201--1207},
  year={2021},
  publisher={Nature Publishing Group UK London}
}

@article{yin2024deep,
  title={Deep random forest with ferroelectric analog content addressable memory},
  author={Yin, Xunzhao and M{\"u}ller, Franz and Laguna, Ann Franchesca and Li, Chao and Huang, Qingrong and Shi, Zhiguo and Lederer, Maximilian and Laleni, Nellie and Deng, Shan and Zhao, Zijian and others},
  journal={Science advances},
  volume={10},
  number={23},
  pages={eadk8471},
  year={2024},
  publisher={American Association for the Advancement of Science}
}

@article{rakka2023dt2cam,
  title={Dt2cam: A decision tree to content addressable memory framework},
  author={Rakka, Mariam and Fouda, Mohammed E and Kanj, Rouwaida and Kurdahi, Fadi},
  journal={IEEE Transactions on Emerging Topics in Computing},
  volume={11},
  number={3},
  pages={805--810},
  year={2023},
  publisher={IEEE}
}

@inproceedings{pedretti2022general,
  title={A general tree-based machine learning accelerator with memristive analog CAM},
  author={Pedretti, Giacomo and Serebryakov, Sergey and Strachan, John Paul and Graves, Catherine E},
  booktitle={2022 IEEE International Symposium on Circuits and Systems (ISCAS)},
  pages={220--224},
  year={2022},
  organization={IEEE}
}

@inproceedings{zhao2024noise,
  title={Noise aware finetuning for analog non-linear dot product engine},
  author={Zhao, Lei and Buonanno, Luca and Natarajan, Aishwarya and Ignowski, Jim and Pedretti, Giacomo},
  booktitle={NeurIPS 2024 Workshop Machine Learning with new Compute Paradigms},
  year={2024}
}

@article{jung2026hybrid,
  title={A Hybrid Digital-Analog Compute-in-Memory Using Content-Addressable Memory With Flexible Multi-Bit Slicing},
  author={Jung, Sangwoo and Lee, Hojin and Park, Jiyong and Lee, Yejin and Park, Dahoon and Shin, Hyunseob and Yoon, Jong-Hyeok and Kung, Jaeha},
  journal={IEEE Transactions on Circuits and Systems I: Regular Papers},
  year={2026},
  publisher={IEEE}
}

@article{wen2026trustworthy,
  author = {Wen, Bo and Gao, Guoyun and Xu, Zhicheng and Jiang, Mingrui and Mao, Ruibin and Qi, Xiaojuan and Chen, Jiezhi and Yin, Xunzhao and Hu, X Sharon and Li, Can},
  journal = {Nature Communications},
  title = {Trustworthy Tree-based Machine Learning by MoS2 Flash-based Analog CAM with Inherent Soft Boundaries},
volume={17},
year = {2026},
}

@inproceedings{qiu2025monolithic,
  title={Monolithic 3D Integration of MoS 2 eDRAM and RRAM for Analog In-Memory Attention Computing},
  author={Qiu, Rui and Gao, Guoyun and Du, Zhiyuan and Yang, Ni and Li, Zefan and Zuo, Dongsheng and Tong, Yunwei and He, Peiyi and Zhang, Hangming and Jiang, Hao and others},
  booktitle={2025 IEEE International Electron Devices Meeting (IEDM)},
  pages={1--4},
  year={2025},
  organization={IEEE}
}

@inproceedings{hanhan2022edam,
  title={EDAM: Edit distance tolerant approximate matching content addressable memory},
  author={Hanhan, Robert and Garz{\'o}n, Esteban and Jahshan, Zuher and Teman, Adam and Lanuzza, Marco and Yavits, Leonid},
  booktitle={Proceedings of the 49th Annual International Symposium on Computer Architecture},
  pages={495--507},
  year={2022}
}

@inproceedings{zhong2023asmcap,
  title={Asmcap: An approximate string matching accelerator for genome sequence analysis based on capacitive content addressable memory},
  author={Zhong, Hongtao and Chen, Zhonghao and Huangfu, Wenqin and Wang, Chen and Xu, Yixin and Wang, Tianyi and Yu, Yao and Liu, Yongpan and Narayanan, Vijaykrishnan and Yang, Huazhong and others},
  booktitle={2023 60th ACM/IEEE Design Automation Conference (DAC)},
  pages={1--6},
  year={2023},
  organization={IEEE}
}

@inproceedings{jahshan2023dash,
  title={DASH-CAM: Dynamic approximate SearcH content addressable memory for genome classification},
  author={Jahshan, Zuher and Merlin, Itay and Garz{\'o}n, Esteban and Yavits, Leonid},
  booktitle={Proceedings of the 56th Annual IEEE/ACM International Symposium on Microarchitecture},
  pages={1453--1465},
  year={2023}
}

@inproceedings{he2024shiftcam,
  title={ShiftCAM: A Time-Domain Content Addressable Memory Utilizing Shifted Hamming Distance for Robust Genome Analysis},
  author={He, Peiyi and Mao, Ruibin and Shan, Keyi and Tong, Yunwei and Xu, Zhicheng and Peng, Muyuan and Luo, Ruibang and Li, Can},
  booktitle={Proceedings of the 43rd IEEE/ACM International Conference on Computer-Aided Design},
  pages={1--9},
  year={2024}
}

@article{jiang2026dshd,
  title={DSHD-CAM: High-Throughput RRAM CAM Leveraging Dynamic Shifted Hamming Distance for Genome Analysis},
  author={Jiang, Chenxin and Zhang, Shen and Zheng, Tiankuo and Li, Rui and Shu, Yuhao and Ha, Yajun},
  journal={IEEE Transactions on Very Large Scale Integration (VLSI) Systems},
  year={2026},
  publisher={IEEE}
}

@inproceedings{laguna2020seed,
  title={Seed-and-vote based in-memory accelerator for dna read mapping},
  author={Laguna, Ann Franchesca and Gamaarachchi, Hasindu and Yin, Xunzhao and Niemier, Michael and Parameswaran, Sri and Hu, X Sharon},
  booktitle={Proceedings of the 39th International Conference on Computer-Aided Design},
  pages={1--9},
  year={2020}
}

@article{liu2023reconfigurable,
  title={A reconfigurable fefet content addressable memory for multi-state hamming distance},
  author={Liu, Liu and Laguna, Ann Franchesca and Rajaei, Ramin and Sharifi, Mohammad Mehdi and Kazemi, Arman and Yin, Xunzhao and Niemier, Michael and Hu, Xiaobo Sharon},
  journal={IEEE Transactions on Circuits and Systems I: Regular Papers},
  volume={70},
  number={6},
  pages={2356--2369},
  year={2023},
  publisher={IEEE}
}

@article{graves2020memory,
  title={In-memory computing with memristor content addressable memories for pattern matching},
  author={Graves, Catherine E and Li, Can and Sheng, Xia and Miller, Darrin and Ignowski, Jim and Kiyama, Lennie and Strachan, John Paul},
  journal={Advanced Materials},
  volume={32},
  number={37},
  pages={2003437},
  year={2020},
  publisher={Wiley Online Library}
}

@article{he2025real,
  title={Real-time raw signal genomic analysis using fully integrated memristor hardware},
  author={He, Peiyi and Wang, Shengbo and Mao, Ruibin and Jiang, Mingrui and Siegel, Sebastian and Pedretti, Giacomo and Ignowski, Jim and Strachan, John Paul and Luo, Ruibang and Li, Can},
  journal={Nature Computational Science},
  volume={5},
  number={10},
  pages={940--951},
  year={2025},
  publisher={Nature Publishing Group US New York}
}

@inproceedings{morris2026np,
  title={NP-CAM: Efficient and Scalable DNA Classification using a NoC-Partitioned CAM Architecture},
  author={Morris, Benjamin F and Molom-Ochir, Tergel and Zhou, Changchun and Chen, Yiran and Jones, Alex K and Li, Hai},
  booktitle={2026 IEEE International Symposium on High Performance Computer Architecture (HPCA)},
  pages={1--14},
  year={2026},
  organization={IEEE}
}

@inproceedings{ghiasi2024megis,
  title={MegIS: High-performance, energy-efficient, and low-cost metagenomic analysis with in-storage processing},
  author={Ghiasi, Nika Mansouri and Sadrosadati, Mohammad and Mustafa, Harun and Gollwitzer, Arvid and Firtina, Can and Eudine, Julien and Mao, Haiyu and Lindegger, Jo{\"e}l and Cavlak, Meryem Banu and Alser, Mohammed and others},
  booktitle={2024 ACM/IEEE 51st Annual International Symposium on Computer Architecture (ISCA)},
  pages={660--677},
  year={2024},
  organization={IEEE}
}

@misc{vemulapati2022orbbasedslamacceleratorsoc,
      title={ORB-based SLAM accelerator on SoC FPGA}, 
      author={Vibhakar Vemulapati and Deming Chen},
      year={2022},
      eprint={2207.08405},
      archivePrefix={arXiv},
      primaryClass={eess.IV},
      url={https://arxiv.org/abs/2207.08405}, 
}

@misc{wu2022memorizingtransformers,
      title={Memorizing Transformers}, 
      author={Yuhuai Wu and Markus N. Rabe and DeLesley Hutchins and Christian Szegedy},
      year={2022},
      eprint={2203.08913},
      archivePrefix={arXiv},
      primaryClass={cs.LG},
      url={https://arxiv.org/abs/2203.08913}, 
}

@misc{yuan2025nativesparseattentionhardwarealigned,
      title={Native Sparse Attention: Hardware-Aligned and Natively Trainable Sparse Attention}, 
      author={Jingyang Yuan and Huazuo Gao and Damai Dai and Junyu Luo and Liang Zhao and Zhengyan Zhang and Zhenda Xie and Y. X. Wei and Lean Wang and Zhiping Xiao and Yuqing Wang and Chong Ruan and Ming Zhang and Wenfeng Liang and Wangding Zeng},
      year={2025},
      eprint={2502.11089},
      archivePrefix={arXiv},
      primaryClass={cs.CL},
      url={https://arxiv.org/abs/2502.11089}, 
}

@misc{lansner2025benchmarkinghebbianlearningrules,
      title={Benchmarking Hebbian learning rules for associative memory}, 
      author={Anders Lansner and Naresh B Ravichandran and Pawel Herman},
      year={2025},
      eprint={2401.00335},
      archivePrefix={arXiv},
      primaryClass={cs.NE},
      url={https://arxiv.org/abs/2401.00335}, 
}

@article{Multiplexed_gradient,
    author = {McCaughan, Adam N. and Oripov, Bakhrom G. and Ganesh, Natesh and Nam, Sae Woo and Dienstfrey, Andrew and Buckley, Sonia M.},
    title = {Multiplexed gradient descent: Fast online training of modern datasets on hardware neural networks without backpropagation},
    journal = {APL Machine Learning},
    volume = {1},
    number = {2},
    pages = {026118},
    year = {2023},
    month = {06},
    issn = {2770-9019},
    doi = {10.1063/5.0157645},
    url = {https://doi.org/10.1063/5.0157645},
    eprint = {https://pubs.aip.org/aip/aml/article-pdf/doi/10.1063/5.0157645/18017061/026118_1_5.0157645.pdf},
}

@misc{deepseekai2025deepseekv32pushingfrontieropen,
      title={DeepSeek-V3.2: Pushing the Frontier of Open Large Language Models}, 
      author={DeepSeek-AI and Aixin Liu and Aoxue Mei and Bangcai Lin and Bing Xue and Bingxuan Wang and Bingzheng Xu and Bochao Wu and Bowei Zhang and Chaofan Lin and Chen Dong and Chengda Lu and Chenggang Zhao and Chengqi Deng and Chenhao Xu and Chong Ruan and Damai Dai and Daya Guo and Dejian Yang and Deli Chen and Erhang Li and Fangqi Zhou and Fangyun Lin and Fucong Dai and Guangbo Hao and Guanting Chen and Guowei Li and H. Zhang and Hanwei Xu and Hao Li and Haofen Liang and Haoran Wei and Haowei Zhang and Haowen Luo and Haozhe Ji and Honghui Ding and Hongxuan Tang and Huanqi Cao and Huazuo Gao and Hui Qu and Hui Zeng and Jialiang Huang and Jiashi Li and Jiaxin Xu and Jiewen Hu and Jingchang Chen and Jingting Xiang and Jingyang Yuan and Jingyuan Cheng and Jinhua Zhu and Jun Ran and Junguang Jiang and Junjie Qiu and Junlong Li and Junxiao Song and Kai Dong and Kaige Gao and Kang Guan and Kexin Huang and Kexing Zhou and Kezhao Huang and Kuai Yu and Lean Wang and Lecong Zhang and Lei Wang and Liang Zhao and Liangsheng Yin and Lihua Guo and Lingxiao Luo and Linwang Ma and Litong Wang and Liyue Zhang and M. S. Di and M. Y Xu and Mingchuan Zhang and Minghua Zhang and Minghui Tang and Mingxu Zhou and Panpan Huang and Peixin Cong and Peiyi Wang and Qiancheng Wang and Qihao Zhu and Qingyang Li and Qinyu Chen and Qiushi Du and Ruiling Xu and Ruiqi Ge and Ruisong Zhang and Ruizhe Pan and Runji Wang and Runqiu Yin and Runxin Xu and Ruomeng Shen and Ruoyu Zhang and S. H. Liu and Shanghao Lu and Shangyan Zhou and Shanhuang Chen and Shaofei Cai and Shaoyuan Chen and Shengding Hu and Shengyu Liu and Shiqiang Hu and Shirong Ma and Shiyu Wang and Shuiping Yu and Shunfeng Zhou and Shuting Pan and Songyang Zhou and Tao Ni and Tao Yun and Tian Pei and Tian Ye and Tianyuan Yue and Wangding Zeng and Wen Liu and Wenfeng Liang and Wenjie Pang and Wenjing Luo and Wenjun Gao and Wentao Zhang and Xi Gao and Xiangwen Wang and Xiao Bi and Xiaodong Liu and Xiaohan Wang and Xiaokang Chen and Xiaokang Zhang and Xiaotao Nie and Xin Cheng and Xin Liu and Xin Xie and Xingchao Liu and Xingkai Yu and Xingyou Li and Xinyu Yang and Xinyuan Li and Xu Chen and Xuecheng Su and Xuehai Pan and Xuheng Lin and Xuwei Fu and Y. Q. Wang and Yang Zhang and Yanhong Xu and Yanru Ma and Yao Li and Yao Li and Yao Zhao and Yaofeng Sun and Yaohui Wang and Yi Qian and Yi Yu and Yichao Zhang and Yifan Ding and Yifan Shi and Yiliang Xiong and Ying He and Ying Zhou and Yinmin Zhong and Yishi Piao and Yisong Wang and Yixiao Chen and Yixuan Tan and Yixuan Wei and Yiyang Ma and Yiyuan Liu and Yonglun Yang and Yongqiang Guo and Yongtong Wu and Yu Wu and Yuan Cheng and Yuan Ou and Yuanfan Xu and Yuduan Wang and Yue Gong and Yuhan Wu and Yuheng Zou and Yukun Li and Yunfan Xiong and Yuxiang Luo and Yuxiang You and Yuxuan Liu and Yuyang Zhou and Z. F. Wu and Z. Z. Ren and Zehua Zhao and Zehui Ren and Zhangli Sha and Zhe Fu and Zhean Xu and Zhenda Xie and Zhengyan Zhang and Zhewen Hao and Zhibin Gou and Zhicheng Ma and Zhigang Yan and Zhihong Shao and Zhixian Huang and Zhiyu Wu and Zhuoshu Li and Zhuping Zhang and Zian Xu and Zihao Wang and Zihui Gu and Zijia Zhu and Zilin Li and Zipeng Zhang and Ziwei Xie and Ziyi Gao and Zizheng Pan and Zongqing Yao and Bei Feng and Hui Li and J. L. Cai and Jiaqi Ni and Lei Xu and Meng Li and Ning Tian and R. J. Chen and R. L. Jin and S. S. Li and Shuang Zhou and Tianyu Sun and X. Q. Li and Xiangyue Jin and Xiaojin Shen and Xiaosha Chen and Xinnan Song and Xinyi Zhou and Y. X. Zhu and Yanping Huang and Yaohui Li and Yi Zheng and Yuchen Zhu and Yunxian Ma and Zhen Huang and Zhipeng Xu and Zhongyu Zhang and Dongjie Ji and Jian Liang and Jianzhong Guo and Jin Chen and Leyi Xia and Miaojun Wang and Mingming Li and Peng Zhang and Ruyi Chen and Shangmian Sun and Shaoqing Wu and Shengfeng Ye and T. Wang and W. L. Xiao and Wei An and Xianzu Wang and Xiaowen Sun and Xiaoxiang Wang and Ying Tang and Yukun Zha and Zekai Zhang and Zhe Ju and Zhen Zhang and Zihua Qu},
      year={2025},
      eprint={2512.02556},
      archivePrefix={arXiv},
      primaryClass={cs.CL},
      url={https://arxiv.org/abs/2512.02556}, 
}

@inproceedings{Millidge2022UniversalHopfield,
  author    = {Beren Millidge and Tommaso Salvatori and Yuhang Song and Thomas Lukasiewicz and Rafal Bogacz},
  title     = {Universal Hopfield Networks: A General Framework for Single-Shot Associative Memory Models},
  booktitle = {Proceedings of the 39th International Conference on Machine Learning (ICML)},
  series    = {Proceedings of Machine Learning Research},
  volume     = {162},
  pages      = {15561--15583},
  year       = {2022},
  publisher  = {PMLR},
  pmid       = {36751405},
  pmcid      = {PMC7614148}
}

@article{radhakrishnan2020overparameterized,
  title={Overparameterized neural networks implement associative memory},
  author={Radhakrishnan, Adityanarayanan and Belkin, Mikhail and Uhler, Caroline},
  journal={Proceedings of the National Academy of Sciences},
  volume={117},
  number={44},
  pages={27162--27170},
  year={2020},
  publisher={National Academy of Sciences}
}

@article{chandra2025episodic,
  title={Episodic and associative memory from spatial scaffolds in the hippocampus},
  author={Chandra, Sarthak and Sharma, Sugandha and Chaudhuri, Rishidev and Fiete, Ila},
  journal={Nature},
  volume={638},
  number={8051},
  pages={739--751},
  year={2025},
  publisher={Nature Publishing Group UK London}
}

@inproceedings{Farzaneh2024,
    author = {Farzaneh, Hamid and De Lima, Joao Paulo Cardoso and Li, Mengyuan and Khan, Asif Ali and Hu, Xiaobo Sharon and Castrillon, Jeronimo},
    title = {C4CAM: A Compiler for CAM-based In-memory Accelerators},
    year = {2024},
    isbn = {9798400703867},
    publisher = {Association for Computing Machinery},
    address = {New York, NY, USA},
    url = {https://doi.org/10.1145/3620666.3651386},
    doi = {10.1145/3620666.3651386},
    booktitle = {Proceedings of the 29th ACM International Conference on Architectural Support for Programming Languages and Operating Systems, Volume 3},
    pages = {164–177},
    numpages = {14},
    location = {La Jolla, CA, USA},
    series = {ASPLOS '24}
    }

@article{Ni2019,
  author    = {Kai Ni and Xunzhao Yin and Ann Franchesca Laguna and Siddharth Joshi and Stefan D{\"u}nkel and Martin Trentzsch and Johannes M{\"u}ller and Sven Beyer and Michael Niemier and Xiaobo Sharon Hu and Suman Datta},
  title     = {Ferroelectric Ternary Content-Addressable Memory for One-Shot Learning},
  journal   = {Nature Electronics},
  year      = {2019},
  volume    = {2},
  number    = {11},
  pages     = {521--529},
  month     = nov,
  doi       = {10.1038/s41928-019-0321-3},
  url       = {https://doi.org/10.1038/s41928-019-0321-3},
  issn      = {2520-1131}
}

@ARTICLE{592312,
    author = {Patterson, David and Anderson, Thomas and Cardwell, Neal and Fromm, Richard and Keeton, Kimberly and Kozyrakis, Christoforos and Thomas, Randi and Yelick, Katherine},
    title = {A Case for Intelligent RAM},
    year = {1997},
    issue_date = {March 1997},
    publisher = {IEEE Computer Society Press},
    address = {Washington, DC, USA},
    volume = {17},
    number = {2},
    issn = {0272-1732},
    url = {https://doi.org/10.1109/40.592312},
    doi = {10.1109/40.592312},
    journal = {IEEE Micro},
    month = mar,
    pages = {34–44},
    numpages = {11}
    }

@misc{dalm2026nodeperturbationeffectivelytrain,
      title={Node Perturbation Can Effectively Train Multi-Layer Neural Networks}, 
      author={Sander Dalm and Marcel van Gerven and Nasir Ahmad},
      year={2026},
      eprint={2310.00965},
      archivePrefix={arXiv},
      primaryClass={cs.LG},
      url={https://arxiv.org/abs/2310.00965}, 
}

@article{Ryu2025,
author = {Ryu, Minjeong and Woo, Jae Seung and Kim, Yeonwoo and Jeon, Joo Hyeon and Cho, Sung In and Choi, Woo Young},
title = {Time-Domain Content-Addressable Memory Based on Single Ambipolar Ferroelectric Memcapacitor for High-Density and Highly-Precise Distance Function Computation},
journal = {Advanced Electronic Materials},
volume = {11},
number = {20},
pages = {e00421},
doi = {https://doi.org/10.1002/aelm.202500421},
url = {https://advanced.onlinelibrary.wiley.com/doi/abs/10.1002/aelm.202500421},
eprint = {https://advanced.onlinelibrary.wiley.com/doi/pdf/10.1002/aelm.202500421},
year = {2025}
}

@article{Boon_gad_desc_homodyne2025,
	author = {Boon, Marcus N. and Cassola, Lorenzo and Ruiz Euler, Hans-Christian and Chen, Tao and van de Ven, Bram and Alegre Ibarra, Unai and Bobbert, Peter A. and van der Wiel, Wilfred G.},
	date = {2025/11/21},
	doi = {10.1038/s41467-025-65155-7},
	id = {Boon2025},
	isbn = {2041-1723},
	journal = {Nature Communications},
	number = {1},
	pages = {10272},
	title = {Gradient descent in materia through homodyne gradient extraction},
	url = {https://doi.org/10.1038/s41467-025-65155-7},
	volume = {16},
	year = {2025}}

@ARTICLE{9921397,
  author={Kleyko, Denis and Davies, Mike and Frady, Edward Paxon and Kanerva, Pentti and Kent, Spencer J. and Olshausen, Bruno A. and Osipov, Evgeny and Rabaey, Jan M. and Rachkovskij, Dmitri A. and Rahimi, Abbas and Sommer, Friedrich T.},
  journal={Proceedings of the IEEE}, 
  title={Vector Symbolic Architectures as a Computing Framework for Emerging Hardware}, 
  year={2022},
  volume={110},
  number={10},
  pages={1538-1571},
  doi={10.1109/JPROC.2022.3209104}}

@article{Mao2022MemristiveMANN,
  author  = {Mao, Rui and Wen, Bo and Kazemi, Amir and He, Peiyi and Li, Can and Strachan, John Paul},
  title   = {Experimentally Validated Memristive Memory Augmented Neural Network with Efficient Hashing and Similarity Search},
  journal = {Nature Communications},
  volume  = {13},
  pages   = {6284},
  year    = {2022},
  month   = oct,
  doi     = {10.1038/s41467-022-33629-7},
  url     = {https://doi.org/10.1038/s41467-022-33629-7}
}

@inproceedings{ni2024tap,
  title={Tap-cam: A tunable approximate matching engine based on ferroelectric content addressable memory},
  author={Ni, Chenyu and Chen, Sijie and Liu, Che-Kai and Liu, Liu and Imani, Mohsen and K{\"a}mpfe, Thomas and Ni, Kai and Niemier, Michael and Hu, Xiaobo Sharon and Zhuo, Cheng and others},
  booktitle={Proceedings of the 43rd IEEE/ACM International Conference on Computer-Aided Design},
  pages={1--9},
  year={2024}
}
    \end{document}